\documentclass[12pt,preprint]{aastex}
\shortauthors{}
\shorttitle{}

\usepackage{subfigure}

\newcommand{\fexii}{Fe~{\sc xii}~195.12~\AA}
\newcommand{\fexiii}{Fe~{\sc xiii}~202.04~\AA}

\begin{document}

\title{Multiple Component Outflows in an Active Region Observed with
the EUV Imaging Spectrometer on {\it Hinode}}       

\author{P. Bryans,\altaffilmark{1,2} 
P. R. Young,\altaffilmark{1,2} \&
G. A. Doschek\altaffilmark{1}}
\altaffiltext{1}{Naval Research Laboratory, 4555 Overlook Ave.\ SW, Washington DC 20375}
\altaffiltext{2}{George Mason University, 4400 University Dr., Fairfax VA 22030}

\begin{abstract} 
We have used the Extreme Ultraviolet Imaging Spectrometer (EIS) on the {\it
Hinode} spacecraft to observe large areas of outflow near an active region. These
outflows are seen to persist for at least 6 days.  The emission line profiles
suggest that the outflow region is composed of multiple outflowing components, 
Doppler-shifted with respect to each other.  We have modeled this scenario by
imposing a double-Gaussian fit to the line profiles. 
These fits represent the
profile markedly better than a single Gaussian fit for Fe~{\sc xii} and {\sc
xiii} emission lines.
For the fastest outflowing components, we find velocities as high as
200~km~s$^{-1}$. However, there remains a correlation between the fitted line
velocities and widths, suggesting that the outflows are not fully resolved by the
double-Gaussian fit and that the outflow may be comprised of further components. 
\end{abstract}

\section{Introduction}
\label{sec:intro}

One of the most significant discoveries of the Extreme Ultraviolet Imaging
Spectrometer (EIS) on the {\it Hinode} spacecraft is the detection of large areas
of outflowing plasma at the boundaries of active regions 
\citep{Dosc07a, Saka07a, Harr08a, Dosc08a}. These outflowing
regions were found to occur in areas of low line emission intensity, often
adjacent to coronal loops. \citet{Dosc07a} also found the
outflowing regions to exhibit larger spectral line widths than found in the much
brighter active region closed loops.  These line widths are in excess of pure
thermal Doppler broadening. \citet{Dosc08a} subsequently found a strong positive
correlation between the outflow velocity (the Doppler shift of the line emission)
and the non-thermal velocity (the width of the emission line).  \citet{Harr07a}
have related the outflowing regions to coronal mass ejections and \citet{Harr08a}
and \citet{Dosc08a} have postulated that the outflowing regions could be 
contributors to the solar wind.  This conclusion is consistent with results based
on completely independent studies of the heliospheric magnetic field, for example
by \citet{Schr03a}.  Given such implications, the further characterization  of
outflowing regions could be important for the understanding of fundamental
physical processes involved in production of the solar wind and mass flow into
the corona.

Emission line widths in excess of their thermal Doppler widths have been
observed since the rocket flight analysis of \citet{Bola75a}.  However, the
origin of the spectral broadening remains unclear; possible explanations
include   turbulence in the atmosphere due to magnetic reconnection
\citep{Park88a} and the presence of coronal waves \citep{Mari92a}.  Earlier
studies of non-thermal line broadening did not link the broadening to bulk mass
flows because there were few high resolution EUV-UV solar spectra coupled to
images such that either the broadening or bulk mass flows could be related to
particular coronal structures.  The outflows under present discussion would have
been difficult to find in earlier studies because of  their appearance in areas
of low intensity in coronal spectral lines, making their measurement difficult. 
It was not until the EIS instrument that the required spatial and spectral
resolution became available to accurately determine the spectral properties of
the outflowing region and marry these with magnetic structures of the solar
atmosphere. Of particular interest among the EIS findings is the correlation of
line shift with line width found by \citet{Dosc08a} and \citet{Hara08a}.   The
fact that those emission lines that show the largest non-thermal Doppler
velocities also display the largest widths raises the possibility that the
outflows may result from multiple flow sites, all Doppler shifted relative to
one another.   Spectral lines at the EIS spatial and spectral resolution might
be convolutions of line emission from multiple unresolved flow sites. This
possibility was suggested by both \citet{Dosc08a}  and \citet{Hara08a} but was
not explored further; their spectral analyses assumed simple Gaussian line
profiles.

An object of the present paper is to determine whether the apparent  excess line
widths can be attributed to the line emission being poorly represented by a
single Gaussian. We expand upon the previous works dealing with outflows in EIS
spectra by attempting to model the outflowing plasma as a blend of outflow sites
with different flow speeds.  This method of analysis depends on the different
flow velocities being sufficiently shifted in relation to one another as to be
spectrally resolved. We thus focus on observations that have previously been
identified as displaying highly asymmetrical line profiles \citep{Dosc08a}. We
also limit our present model to an outflow region that can be well-represented by
a sum of two flow velocities, with the emission line profiles modeled as a sum of
two Gaussian components. The remainder of this paper is organized as follows:  in
Section~\ref{sec:obs} we outline the observations and explain the  EIS data
reduction procedures. In Section~\ref{sec:results} we present the results of the
double Gaussian line fitting technique and discuss possible interpretations in
Section~\ref{sec:discuss}.

\section{Observations and Data Reduction}
\label{sec:obs}

The EIS instrument on {\it Hinode} is described by \citet{Culh07a}; more details
are also given by \citet{Kore06a}. The {\it Hinode} mission is described by
\citet{Kosu07a}.  The EIS observations discussed in the present paper utilize the
1\arcsec\ slit rastered from west to east in 1\arcsec\ increments to build up a
two-dimensional spectral image.  The two observing programs used here raster
areas of $256 \times 256$ and $460 \times 384$ pixels, returning 20 and 24
spectral windows, respectively.

\begin{figure}[htp]
     \centering
     \vspace{-.45in}
     \subfigure{\includegraphics[width=.32\textwidth]{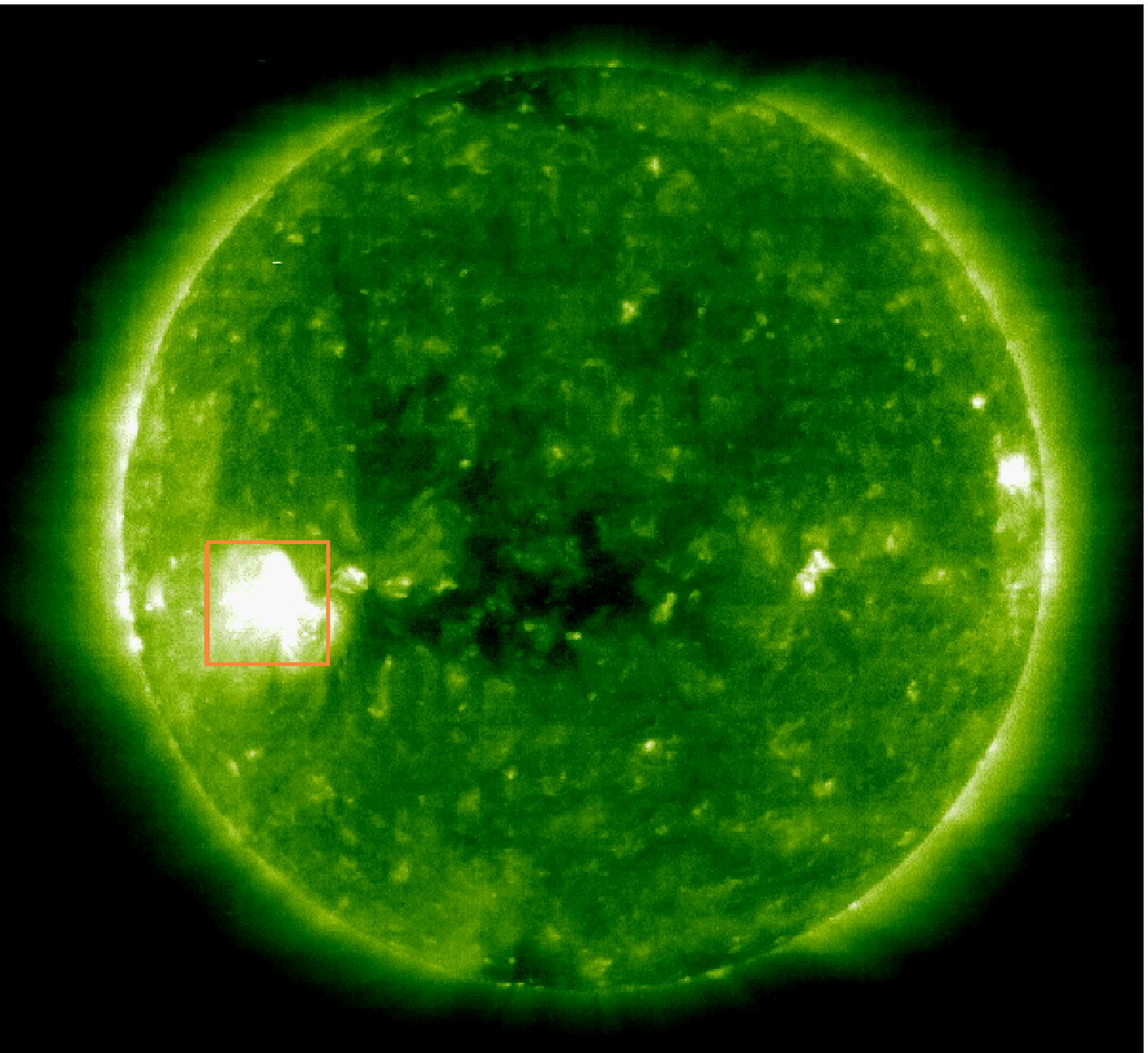}}
     \subfigure{\includegraphics[width=.32\textwidth]{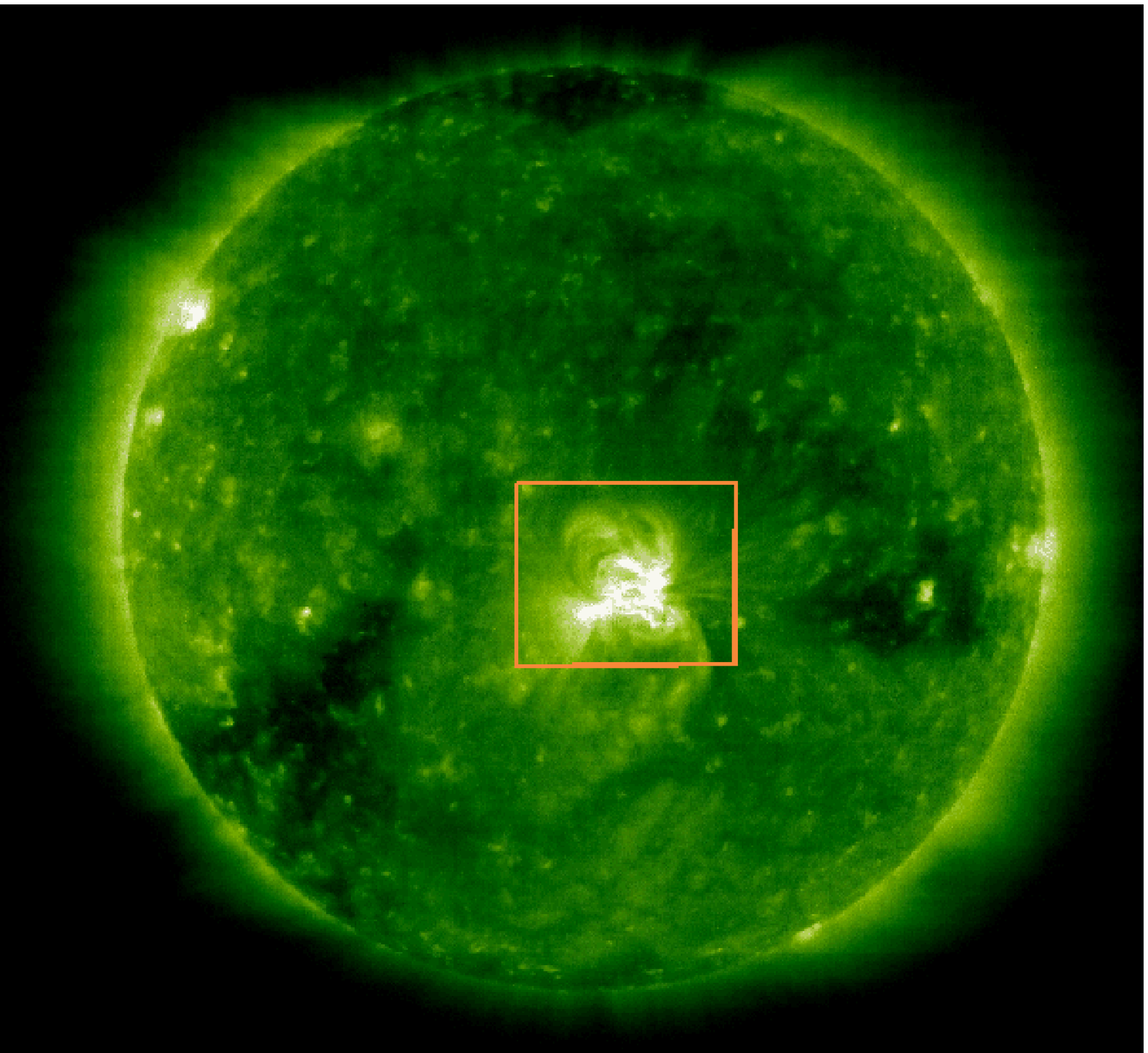}}
     \subfigure{\includegraphics[width=.32\textwidth]{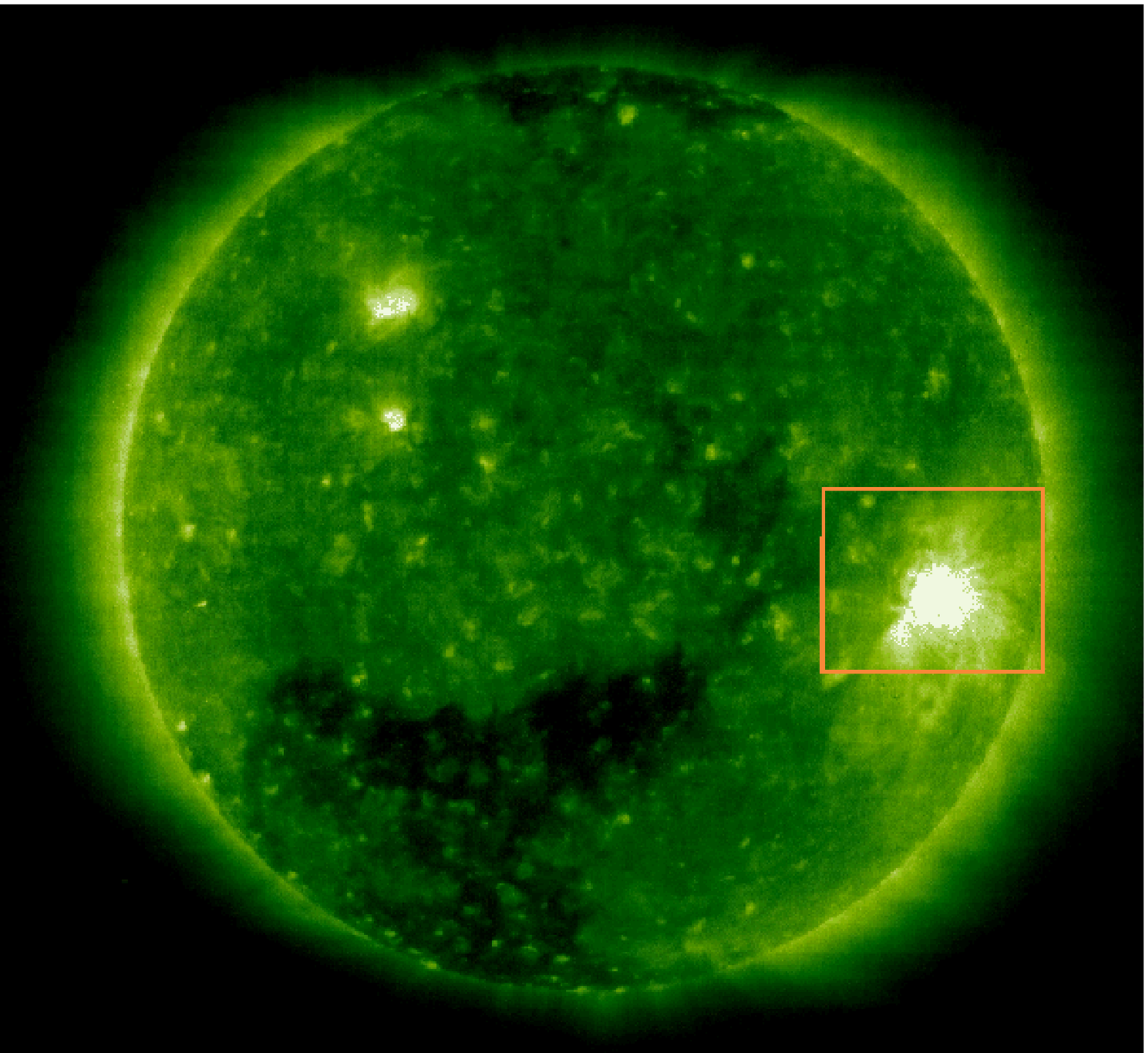}}
     \caption{Full disk images from the 195\AA\ filter of EIT at 
     Dec 09 00:00 (left; corresponding to EIS observation 1),
     Dec 12 11:12 (middle; corresponding to EIS observation 7), and
     Dec 15 18:13 (right; corresponding to EIS observation 10).
     Images courtesy of the SOHO EIT Consortium; 
     SOHO is a joint ESA-NASA program.}
     \label{fig:eit}
\end{figure}

For the purposes of this paper we have analyzed the active region (AR) 10978. 
This AR was previously identified as displaying an outflow region with possible
multiple components by \citet{Dosc08a}.   
\citet{Delz08a} measured outflows from a single active region and found
that they persisted for at least 4 days. We expand on these two works  by
tracking the evolution of the AR observed by \citet{Dosc08a} over 7
days and characterize the outflows through two Gaussian rather than single
Gaussian fits to the emission line profiles.
We use 10 separate EIS raster observations
spanning from 2007 December 09 to 2007 December 15.   To give context of the size
and location of the AR, we show full disk Extreme Ultraviolet Imaging Telescope
(EIT) images at 
Dec 09 00:00,
Dec 12 11:12, and
Dec 15 18:13
in Figure~\ref{fig:eit}.  
The EIS field of view is highlighted as orange boxes in Figure~\ref{fig:eit} 
and the times and locations of
the EIS observations are summarized in Table~\ref{tab:obs}.  The locations are
given in arcseconds relative to Sun center, with positive positions north and
west. Exposure times are for each position in the raster. We label the
observations 1--10 for ease of reference henceforth.

We have reduced the observational data using the standard EIS software data
reduction package eis\_prep as described in the SolarSoft software distribution
\citep{ssw}.  This includes flagging any saturated data as missing, removing the
CCD pedestal and dark current, removing cosmic rays and hot pixels, and
accounting for detector bias. In addition, there exists a misalignment between
the EIS CCD and slit that causes the observed spectra to lie at an angle to the
CCD column.  The magnitude of this `slit tilt' has been estimated by averaging
the velocities over a large raster of the quiet Sun and we use these values to
correct for its effect here. Finally, there is an instrumental effect---due to
the temperature variation over the {\it Hinode} orbit---that causes a
quasi-periodic variation in the line centroid position. We corrected for this
`thermal drift' by selecting a quiet Sun region near but outside of the active
region for each raster observation and assumed that the average of the line
centroids in this region represented the rest wavelength.  This allows an
adjustment of all the wavelengths in the raster that removes the orbital drift.
In some of the observations the South Atlantic Anomaly (SAA) results in
enhanced particle hits on the detector.  This effect occurs in a particular
part of the {\it Hinode} orbit and results in vertical stripes of noise
on the raster images.  These areas have been omitted from the analysis.

The Fe~{\sc xii} emission line at 195.12~\AA\ is the strongest line observed in
the observations analyzed here.  Due to the high count statistics, this line is
commonly used when deriving both thermal and non-thermal plasma velocities from
EIS observations.  However, care must be taken when analyzing this line due to a
blend with another Fe~{\sc xii} line at 195.18~\AA. The ratio of these two lines
depends on electron density.   In addition to resulting in an increase in the
observed line width,  the presence of this blend can also shift the line centroid
longward of the actual wavelength of the 195.12~\AA\ emission, depending on the
density.  The density of the outflow region analyzed here was calculated by
\citet{Dosc08a} to be $\lesssim 10^9$~cm$^{-3}$. At these densities the
195.18~\AA\ line should not significantly affect the 195.12~\AA\ emission 
\citep[see][]{Youn09a}.

There are two other Fe~{\sc xii} lines at similar wavelengths that are not
blended: at 192.39~\AA\ and 193.51~\AA.  These lines have emissivities of the
order of 1/3 and 2/3 that of the 195.12~\AA\ line, respectively. Unfortunately,
the position of the 193.51~\AA\ line on the detector  is contaminated with dust
so a section of the observed data is lost.  For the observations described in
this paper, this missing section happens to be directly aligned with the
outflowing region so a significant portion of the data from the outflow is not
available.  
In addition to the Fe~{\sc xii} lines, we analyze emission from 
Si~{\sc vii}~275.67~\AA,
Fe~{\sc viii}~185.21~\AA, Fe~{\sc x}~184.54~\AA, \fexiii, Fe~{\sc
xiv}~264.79~\AA, Fe~{\sc
xiv}~274.20~\AA, Fe~{\sc xv}~284.16~\AA\
and Fe~{\sc xvi}~262.98~\AA. These emission lines come from ions
with formation temperatures spanning  $4.0\times 10^5$--$2.5\times 10^6$~K
\citep{Brya09a}.

As explained in Section~\ref{sec:intro}, an observation of this AR has previously
been studied by \citet[][observation 4 of the present paper]{Dosc08a} and in
this paper we model the line profiles as the sum of two Gaussians with a linear
background. Figure~\ref{fig:spectra} shows examples of the Fe~{\sc xii}
192.39~\AA\ and 195.12~\AA\ and \fexiii\ spectra for an
individual spatial pixel in the
observation 4 outflow region. Each example shows significant emission shortward
of the peak emission wavelength.  These asymmetric blue wings are typical of the
line emission in this region.

To ensure a robust fit to the line profiles we require that the two Gaussians
have the same width.  While this assumption is not necessarily true of the
emission, we have to impose some restrictions on the fitting because leaving
every parameter to vary freely results in unsatisfactory fits.
Figure~\ref{fig:spectra} shows examples of these fits. We subsequently refer to
the largest (in intensity) Gaussian as the primary component and the smaller,
blue-shifted Gaussian as the secondary component.

\section{Results}
\label{sec:results}

A double-Gaussian fit was performed over each observation region for each of the
observed emission lines.  Figure~\ref{fig:intensity} shows the intensity maps
(from the sum of both Gaussians) of the \fexii\ line for each observation. The
emission from the primary component is dominant, and significant contribution
from the secondary component is only seen in isolated regions, as shown in
Figure~\ref{fig:intensity secondary}.
We show red contours of the secondary component on  Figure~\ref{fig:intensity} 
to indicate the location of the outflows.

Figure~\ref{fig:velocity} shows the line centroid shift of the primary component
of the \fexii\ line for each raster.  Comparison of Figures~\ref{fig:intensity
secondary} and \ref{fig:velocity} show that the outflowing regions are also those
that display a secondary Gaussian component in their emission profiles. It is
also interesting to note the evolution of the outflowing region over time. We
initially observe (9 December) outflowing material directly to the west of the AR
and subsequently observe outflowing regions on both the east and west of the 
AR.  While the western outflow region remains for the entire observing period,
the apparent velocity of the flow is significantly decreased by the final
observation (15 December). While this may suggest that the region of plasma
outflow is changing position with time, it is probable that this is a line-of
sight effect due to the solar rotation. The AR loop system in the early
observations is oriented close to the east limb which would tend  to obscure
low-lying outflowing plasma to the east of the active region (trailing
component). As the Sun rotates, the loops then obscure the western portion
(leading component) of the outflow as the active region approaches the west
limb.  We discuss these projection effects further in Section~\ref{sec:discuss}.
What is clear, however, is that the outflow persists for the entire duration that
the AR is observable.

As an aside, a brief discussion of the apparent red-shift of the AR loops is
warranted.  While Figure~\ref{fig:velocity} appears to suggest downflowing plasma
at the core of the AR, this red-shift is actually due to a blend in the
\fexii\ line.   We previously noted, in Section~\ref{sec:obs}, that this blend is
insignificant at the low density of the outflowing region. However, the higher
density of the core of the AR is sufficient to increase the relative intensity of
the Fe~{\sc xii}~195.18~\AA\ line such that the perceived centroid and width of
the \fexii\ line are altered. A full discussion of this blend can be found in
\citet{Youn09a}, but we note here that the red-shift of the AR core is
a density effect rather than plasma flow.

Comparison of Figures~\ref{fig:intensity secondary} and \ref{fig:velocity} shows
some interesting differences between the western and eastern outflow regions. The
secondary component of the western region, while strongest at the base of the
outflow region, is seen to be present almost entirely throughout the outflowing
region for observations 1 through 8.
However, the same is not true of the eastern region.
This is best illustrated in observations 9 and 10 where one can see extended
areas of outflow in Figure~\ref{fig:velocity}, but only isolated
regions of intensity of the secondary Gaussian component in
Figure~\ref{fig:intensity secondary}.

In Figure~\ref{fig:histogram main} we show histograms of the flow velocity of the
primary Gaussian component at each observation time.  The velocity is calculated
as the shift in line centroid from a reference wavelength of 195.12~\AA, assumed
to be the rest wavelength in constructing Figure~\ref{fig:velocity}. For each
plot in Figure~\ref{fig:histogram main} (representing a different observation) we
show velocity histograms for the western outflow region as black lines and those
for the eastern outflow region as red lines. We have isolated the outflowing
regions from the surrounding plasma by including in these histograms only those
pixels where the intensity of the secondary Gaussian is at least 10\% of the
primary Gaussian. Over the 10 rasters, we find a median velocity in the range
$\sim5-13$~km~s$^{-1}$ for the western outflow region, and $\sim0-10$~km~s$^{-1}$
for the eastern outflow region. For the western outflow region of observation 4,
\citet{Dosc08a} reported maximum flow speeds on the order of 45~km~s$^{-1}$ with
a median shift of $\sim20$~km~s$^{-1}$.  Our results for the flow speed of the
primary Gaussian are slower than the \citet{Dosc08a} values, as one would expect
from fitting a double Gaussian rather than a single to the emission profile.
However, it is interesting to note that the apparent outflow observed from a
single Gaussian fit cannot be entirely explained by the presence of the secondary
Gaussian component; we find that the primary Gaussian is also shifted shortward
of the rest wavelength.

The more interesting result is the velocity derived from the secondary
component.   Histograms of the same format as Figure~\ref{fig:histogram main} are
plotted for the secondary Gaussian in Figure~\ref{fig:histogram minor}. Here we
find extremely large velocities: a median of $\sim100-130$~km~s$^{-1}$  for the
western outflow region and $\sim90-120$~km~s$^{-1}$  for the eastern outflow
region. For both regions, some spatial pixels show velocities as high as 
200~km~s$^{-1}$.

As outlined in Section~\ref{sec:intro}, previous works have found a correlation
between the outflow speed and line width in AR outflows.  We can determine
whether our new double Gaussian fitting resolves this correlation by plotting the
Doppler velocity versus the full width at half maximum (FWHM) of both Gaussian
components. In Figures~\ref{fig:velocity vs width main} and \ref{fig:velocity vs
width minor} we show these relations for the primary and secondary Gaussian 
components respectively. We use black dots to represent the western outflow
region and red dots to represent the eastern outflow region. Fitting two
Gaussians to the spectrum rather than one naturally reduces the width with
respect to a single Gaussian fit.  However, there remains some evidence of a
correlation between the  velocity and width in the primary component. This may
suggest that the double Gaussian fit still does not resolve the outflow region;
we discuss this further in Section~\ref{sec:discuss}.

The results discussed thus far have pertained to the \fexii\ emission.   
For the other Fe~{\sc xii} lines, at 192.39~\AA\
and 193.51~\AA, we find results that match well with those of the \fexii\
emission described above. Results from \fexiii\ are also in accordance with those
of the Fe~{\sc xii} lines.
Low temperature lines from Si~{\sc vii} and Fe~{\sc viii}
were not found to have any significant
asymmetry to the emission profile and could be accurately represented by a
single Gaussian without the need for a secondary blue-shifted component.

The greatest difficulty in analyzing emission from ions spanning a greater
temperature range is finding lines that are free of blends.
The Fe~{\sc x} emission line at 184.54~\AA\ is close
to another line at a slightly shorter wavelength.  Contamination  from this line 
does not allow a satisfactory double Gaussian to be fit to the 184.54~\AA\
emission.  Similarly, emission from  Fe~{\sc xiv}~274.20~\AA\ is blended with a
Si~{\sc vii} line which hinders a double Gaussian fitting.

At higher temperatures, emission from Fe~{\sc xiv}~264.79~\AA,
Fe~{\sc xv}~284.16~\AA\ and Fe~{\sc
xvi}~262.98~\AA\ are not affected by nearby lines. However, while emission from
these ions is strong in the core of the active region, emission in the
outflowing regions is relatively weak. It is thus difficult to satisfactorily
fit a double Gaussian profile to the emission. It is possible to improve the
count statistics of these emission lines by summing adjacent pixels. We have
done this for areas of $5\times 5$ pixels in size, increasing the detected flux
by a factor of 25 at the expense of spatial resolution. 
For these lines we again find no
significant asymmetry to the emission profile. 
We do, however, find a single Gaussian fit to display
outflow velocities of the same order as those determined from the primary
component of the double Gaussian fits to the Fe~{\sc xii} and Fe~{\sc xiii}
lines.

\section{Discussion and Conclusions}
\label{sec:discuss}

\begin{figure}[htp]
     \centering
     \subfigure[Fe~{\sc xii}
     192.39~\AA]{\includegraphics[width=.3\textwidth]{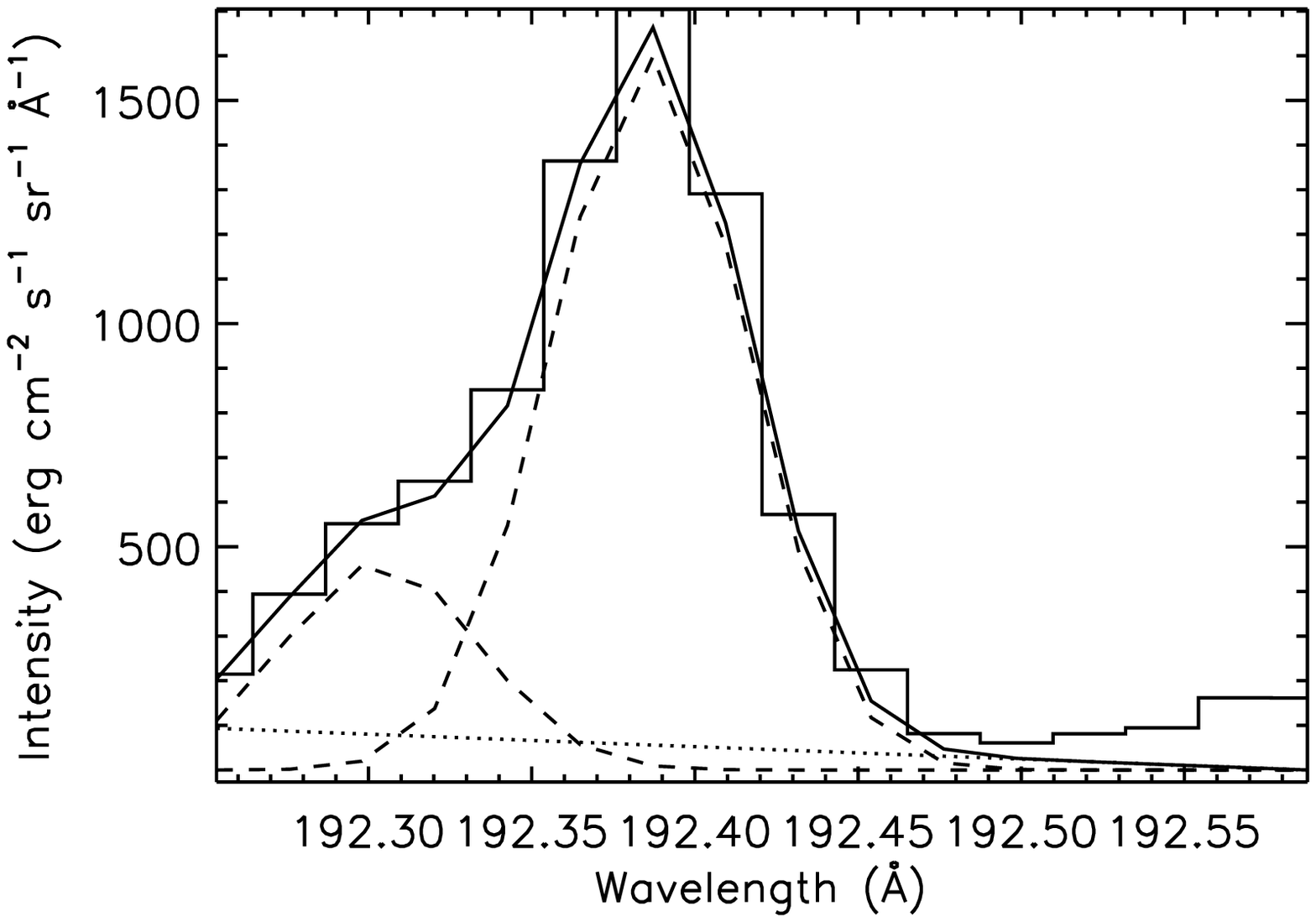}}
     \subfigure[\fexii]{\includegraphics[width=.3\textwidth]{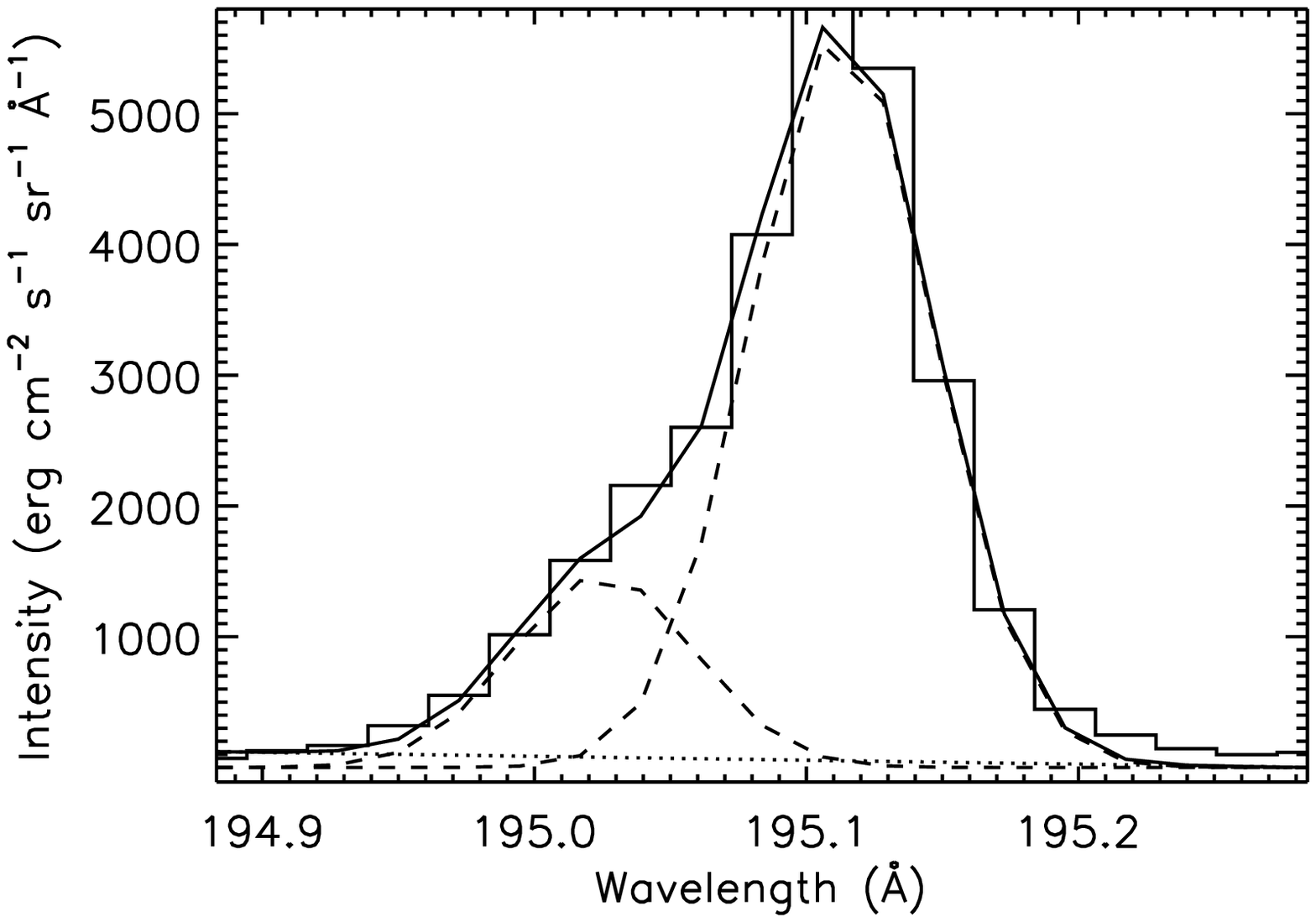}}
     \subfigure[\fexiii]{\includegraphics[width=.3\textwidth]{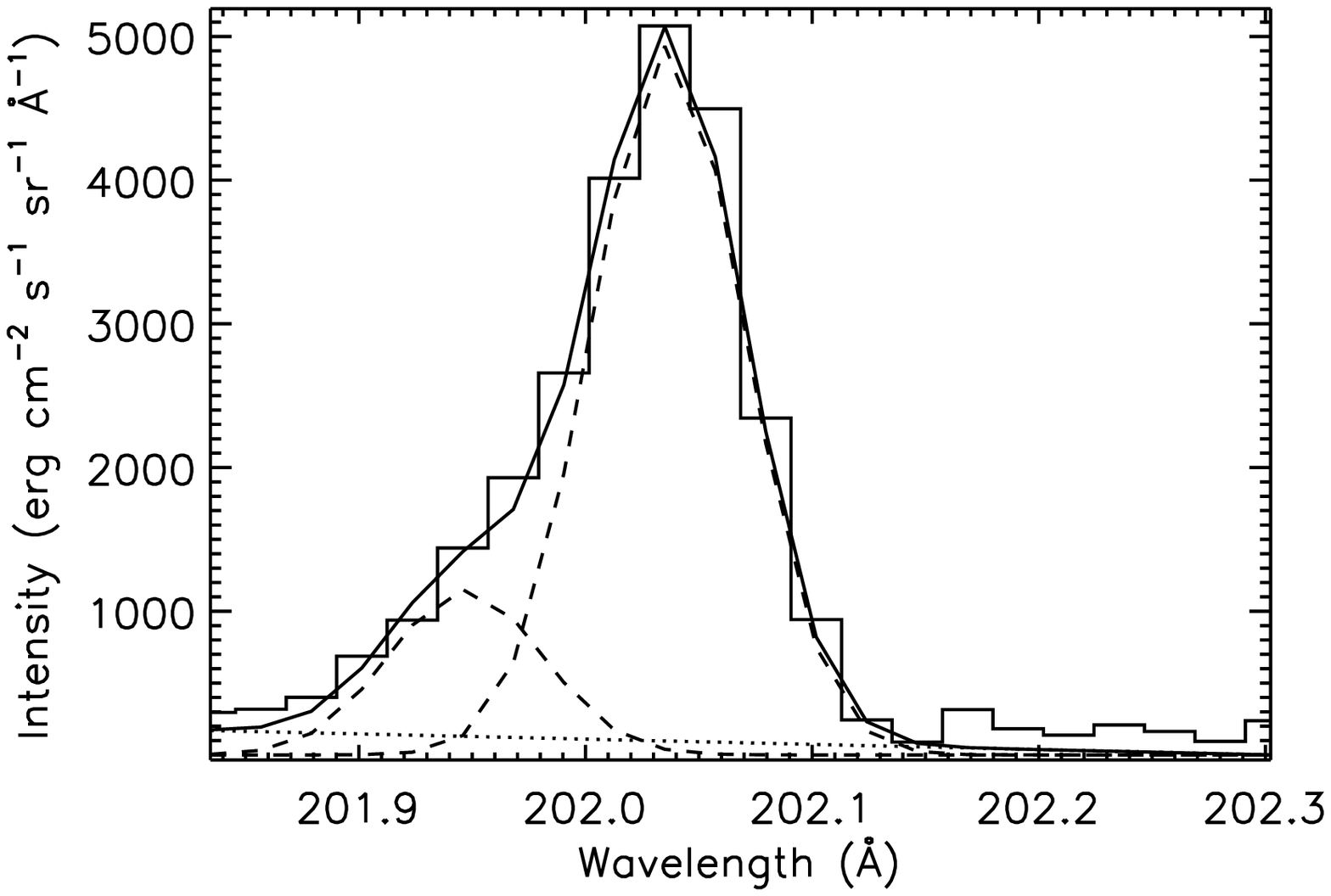}}
     \caption{Observed EIS spectra from a single pixel of the outflow region 
              after data calibration.
	      The chosen pixel is, in arcsec from Sun center, 
	      at (-350,-110) from the observation
	      on 2007 Dec 10 00::19:27.
	      The double Gaussian fit is superimposed (solid curve) with the
	      primary and secondary Gaussian components shown (dashed curves).}
     \label{fig:spectra}
\end{figure}

\begin{figure}[htp]
     \centering
     \vspace{-.45in}
     \subfigure{\includegraphics[width=.40\textwidth]{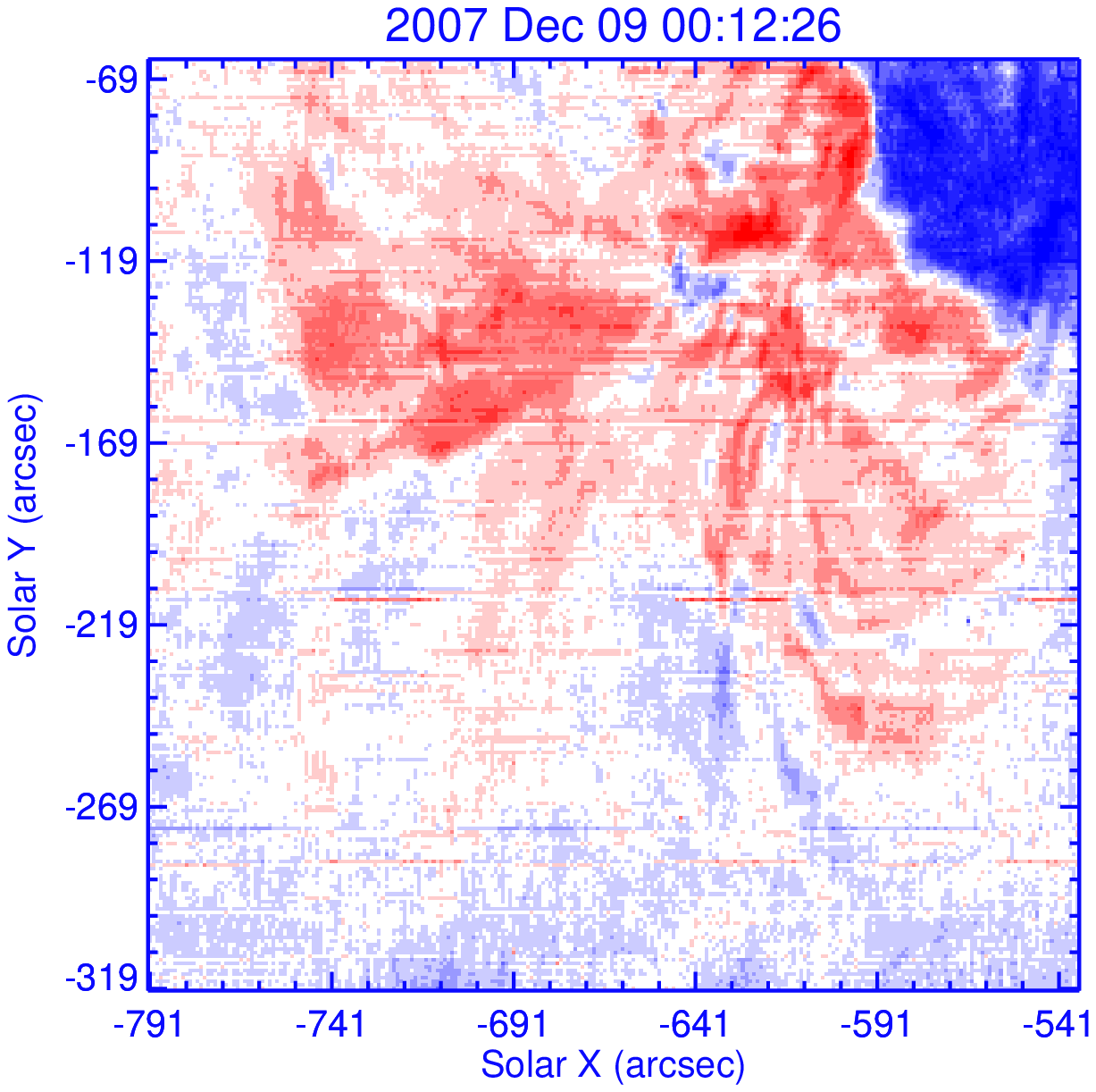}}
     \subfigure{\includegraphics[width=.40\textwidth]{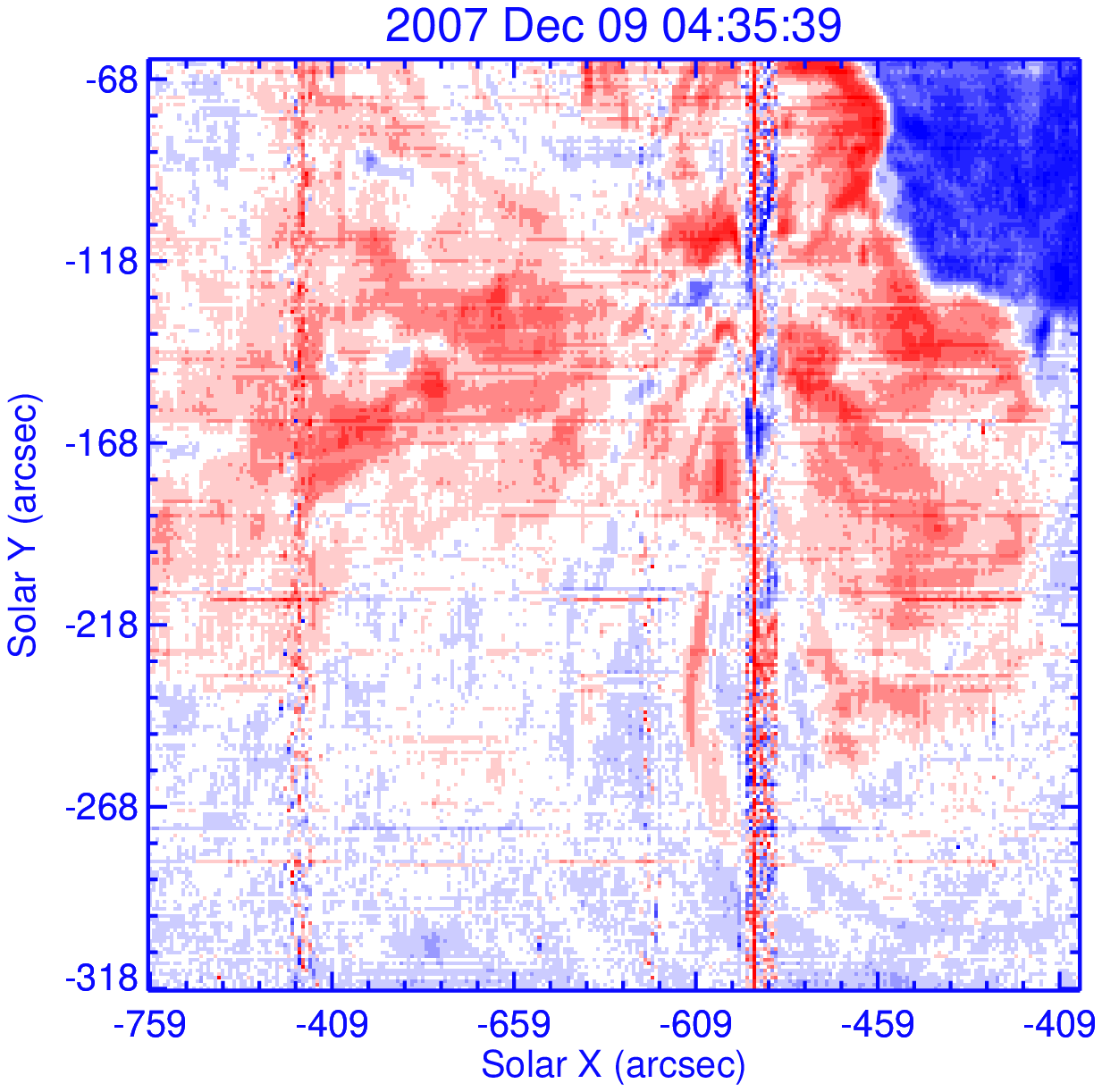}}\\
     \vspace{-.2in}
     \subfigure{\includegraphics[width=.40\textwidth]{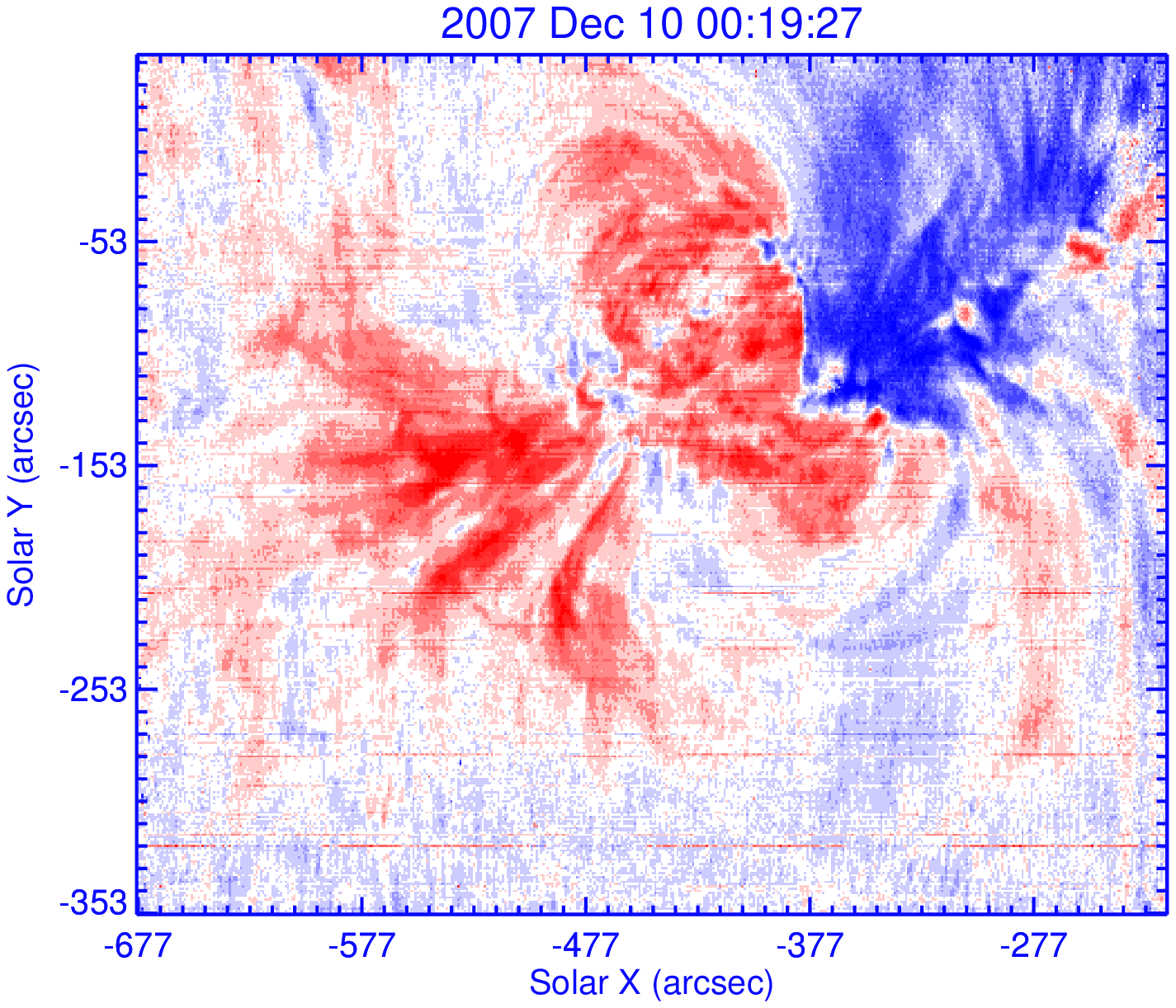}}
     \subfigure{\includegraphics[width=.40\textwidth]{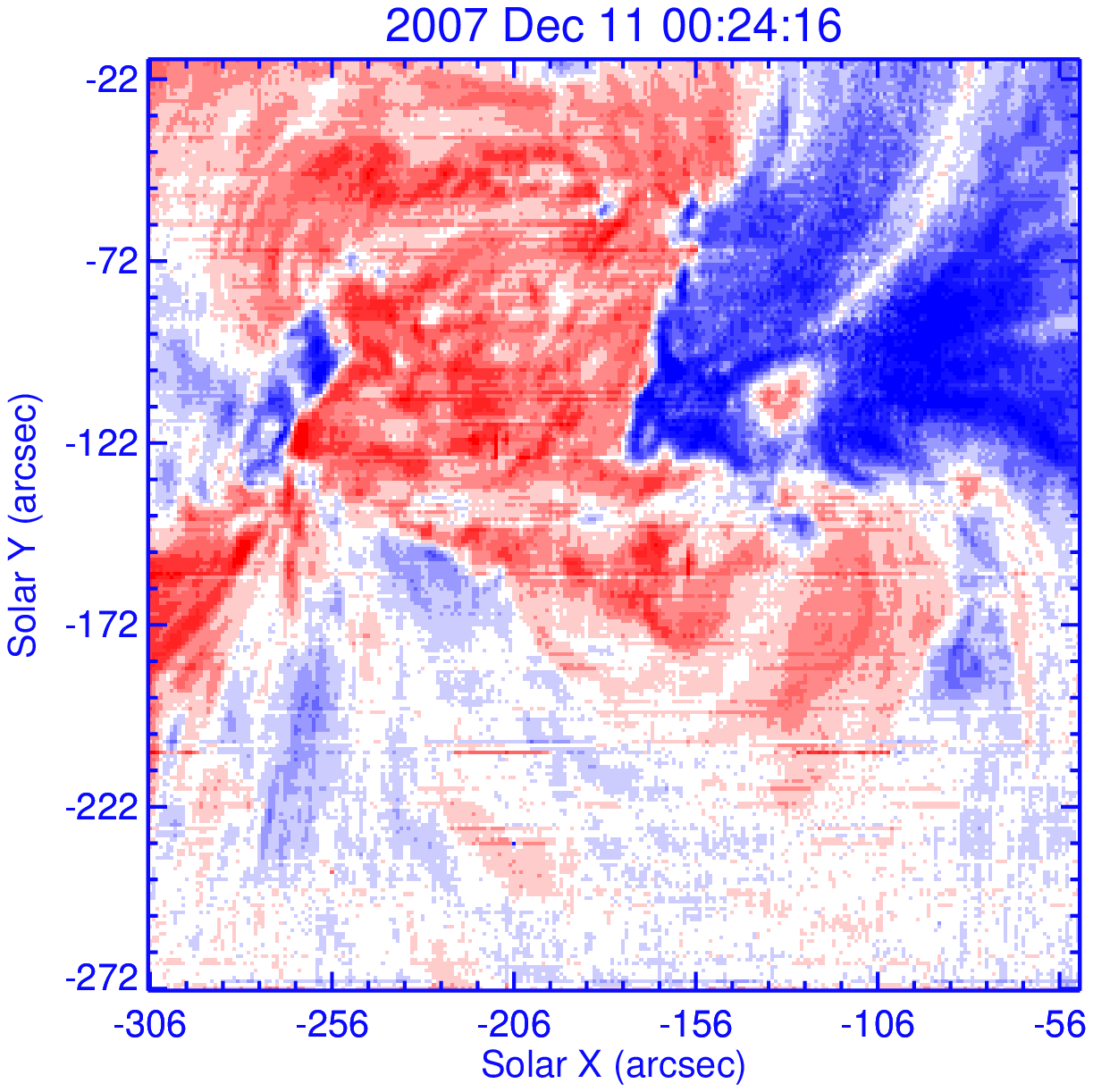}}\\
     \vspace{-.2in}
     \subfigure{\includegraphics[width=.40\textwidth]{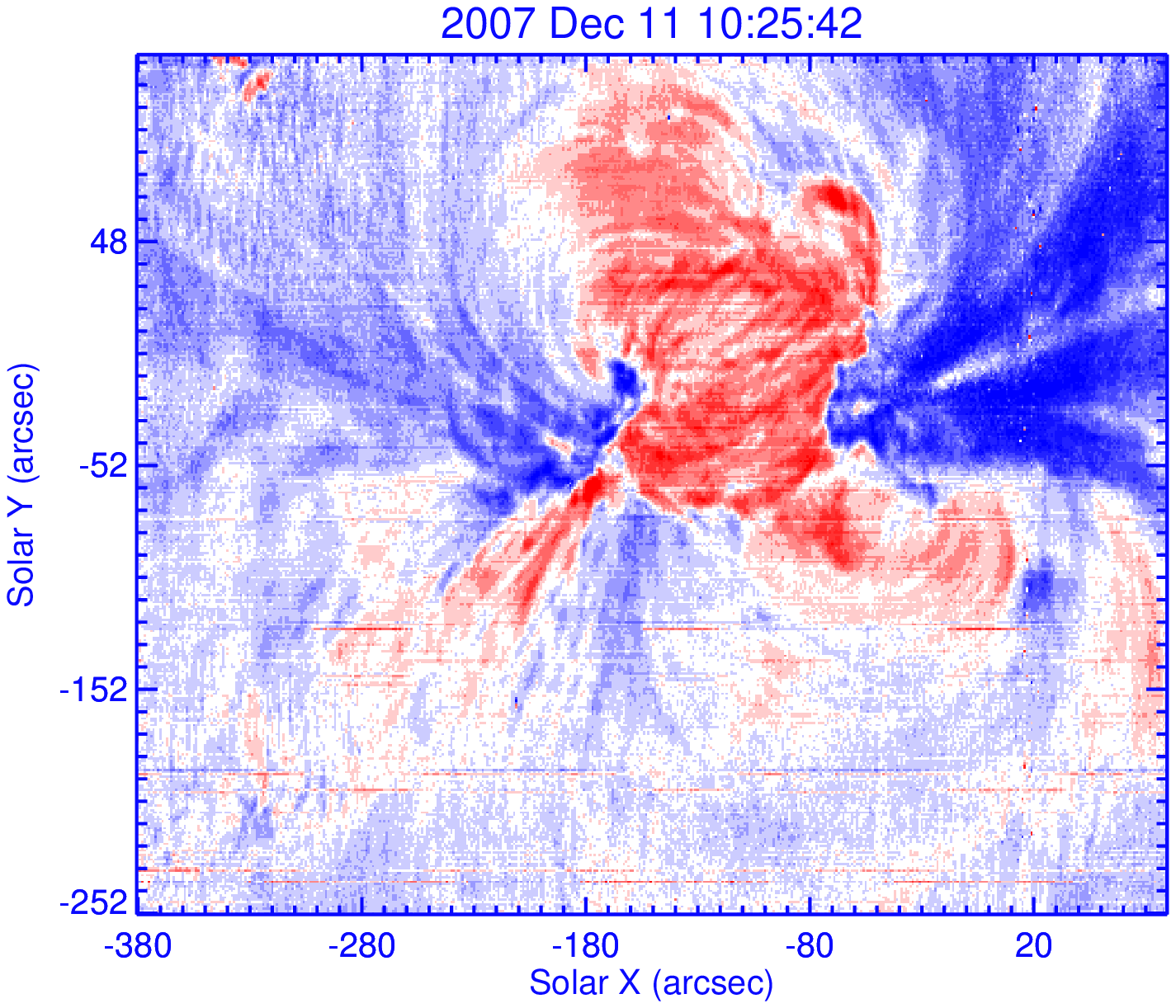}}
     \subfigure{\includegraphics[width=.40\textwidth]{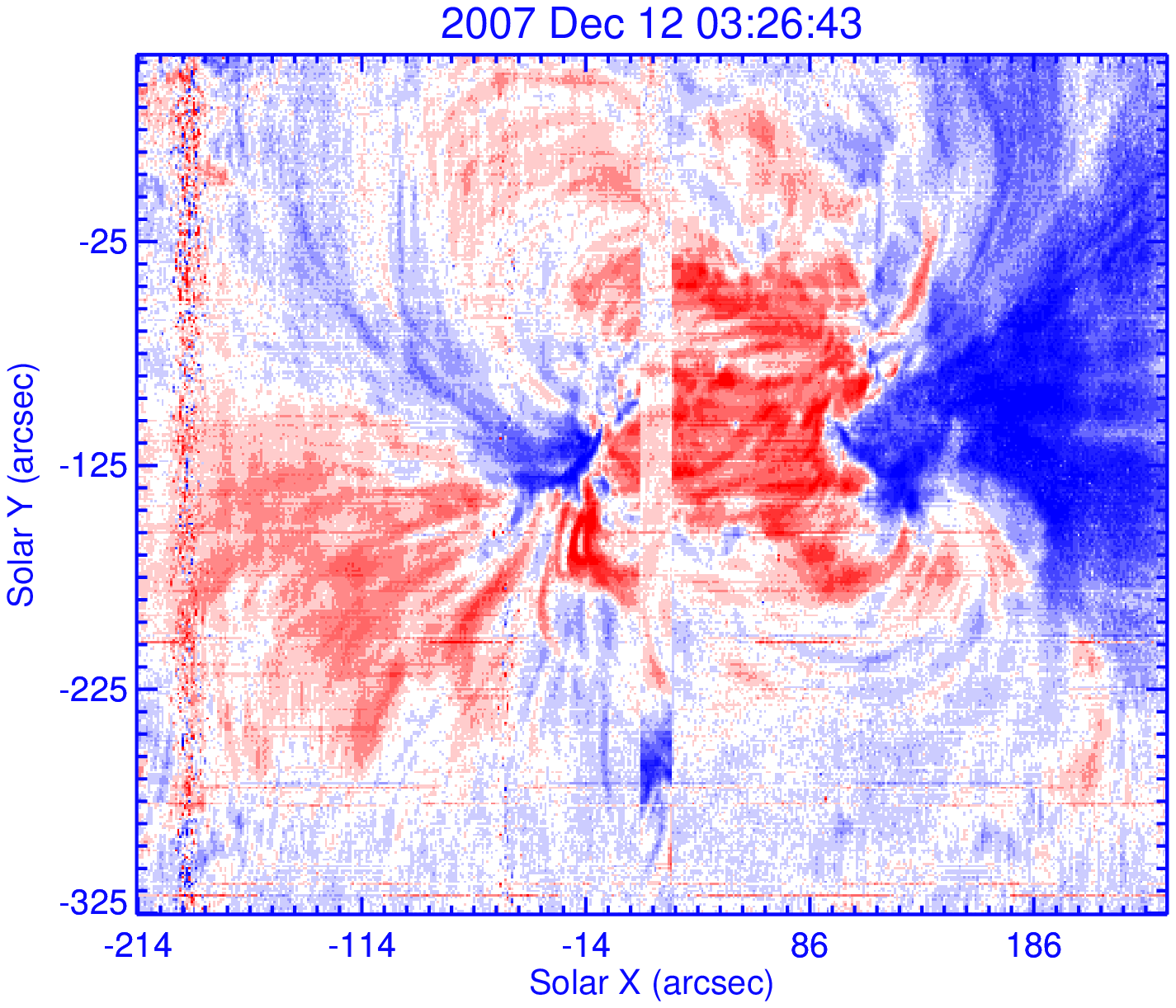}}\\
     \vspace{-.2in}
     \subfigure{\includegraphics[width=.40\textwidth]{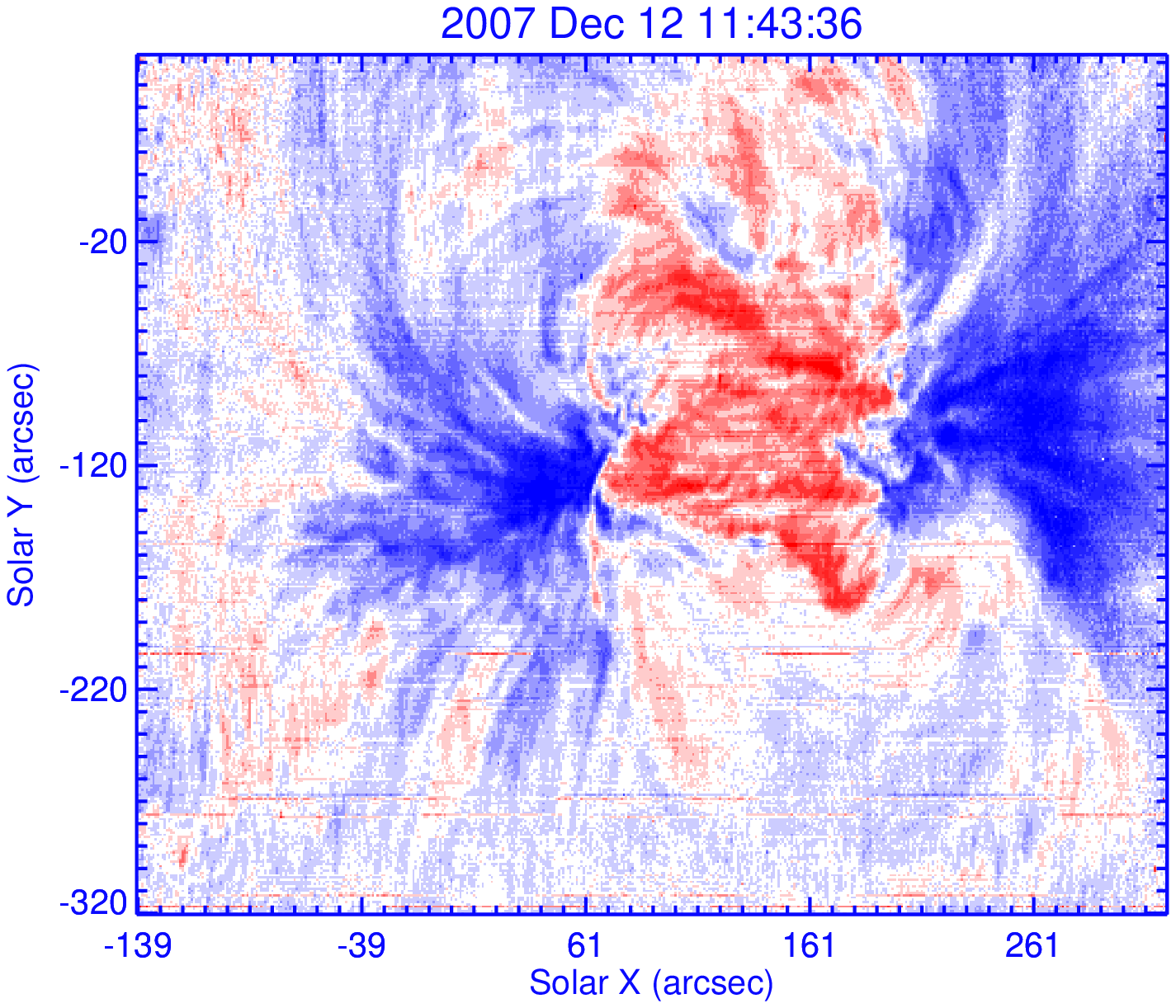}}
     \subfigure{\includegraphics[width=.40\textwidth]{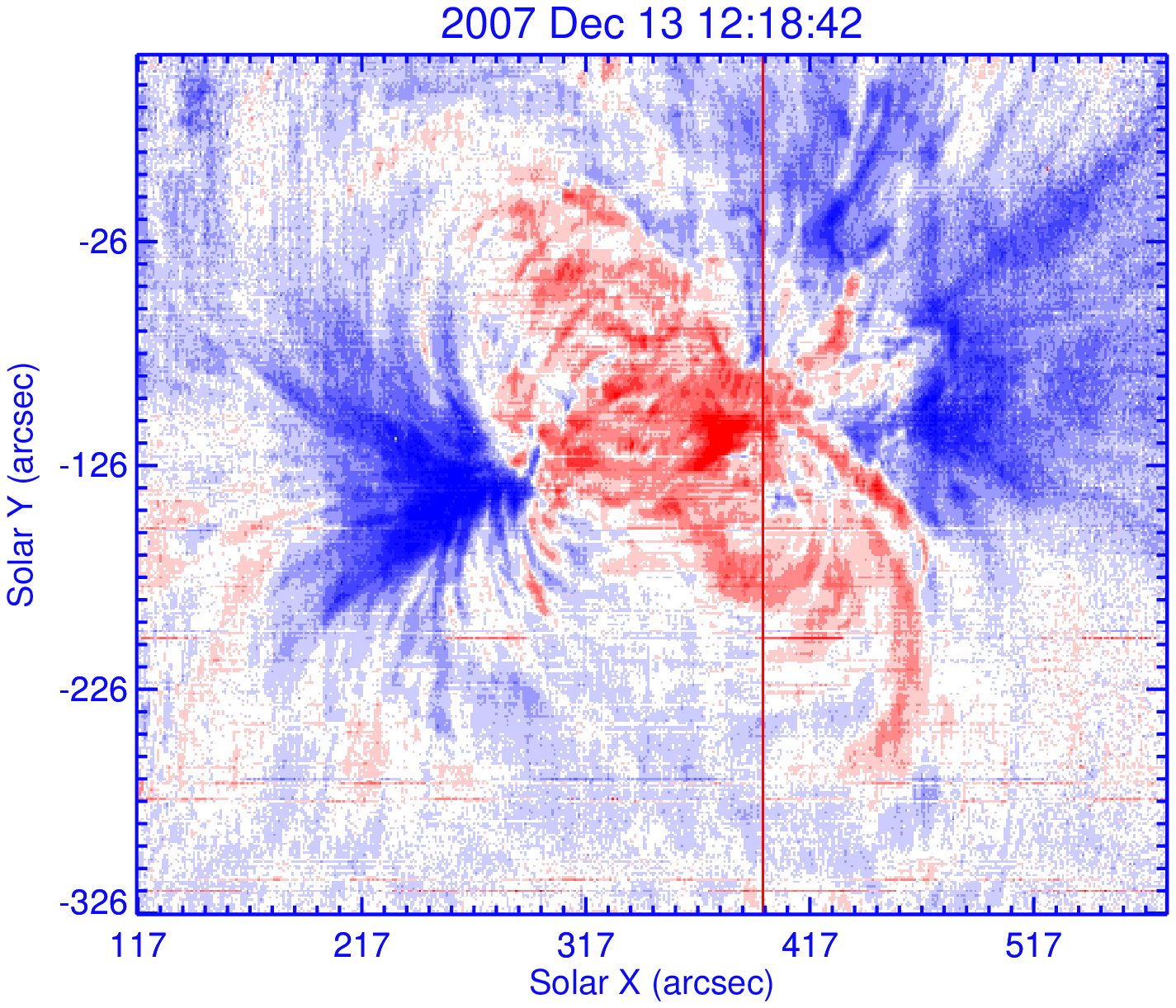}}\\
     \vspace{-.2in}
     \subfigure{\includegraphics[width=.40\textwidth]{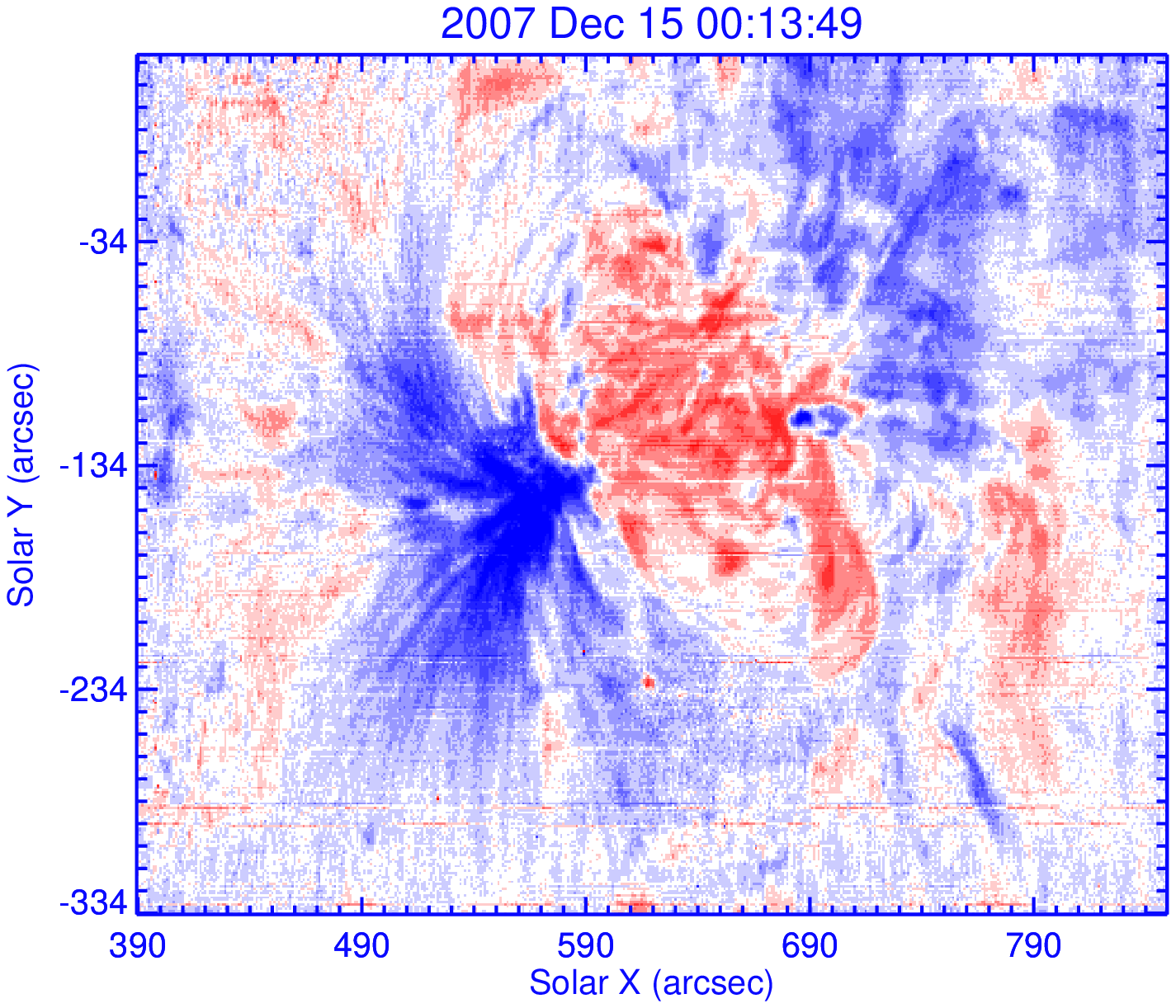}}
     \subfigure{\includegraphics[width=.40\textwidth]{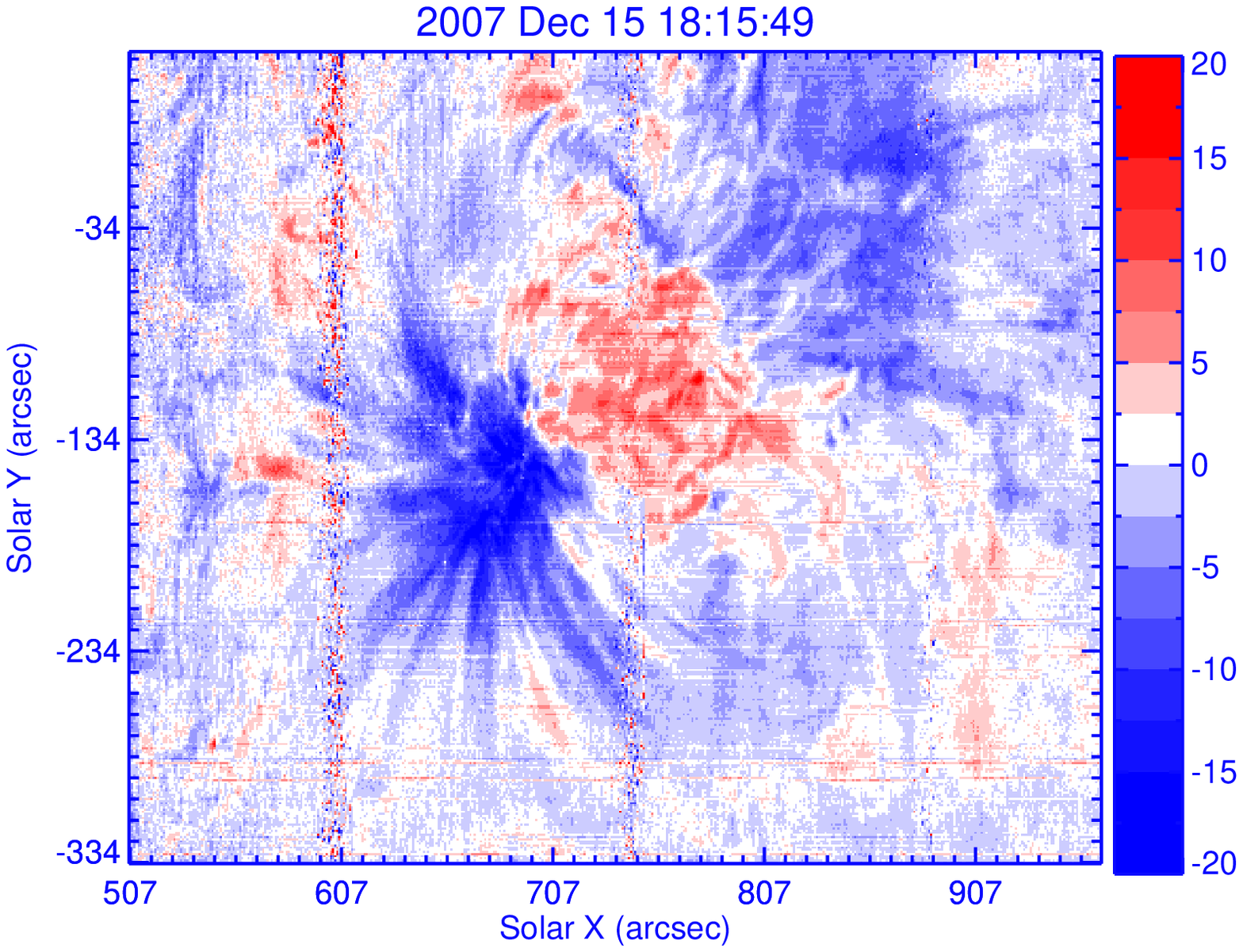}}
     \caption{
     Velocity maps of the primary Gaussian 
     for each of the 10 EIS observations.  Blue represents an upflow and red
     a downflow.}
     \label{fig:velocity}
\end{figure}

As shown in Figure~\ref{fig:spectra}, a double Gaussian representation of the
emission line profiles is a more accurate fit to the data than a single Gaussian
representation for certain emission lines in the outflowing region.
From the results presented in Section~\ref{sec:results}, we see
that the double Gaussian representation is only applicable to the outflowing
regions. Within these regions we find that the outflow is not only due to the
fast moving secondary Gaussian component, but the primary component is also
blue-shifted (see Figure~\ref{fig:velocity}).

By its nature the double Gaussian fit results in a reduction in the Doppler
velocity of the primary component compared to that of a single Gaussian fit.  We
find velocities for this component of the order of 10~km~s$^{-1}$, compared to
$\sim20$~km~s$^{-1}$ found by \citet{Dosc08a} for the same outflow. However, the
secondary component exhibits significantly greater outflow velocities, often as
large as 200~km~s$^{-1}$.

\begin{figure}[htp]
     \centering
     \vspace{-.45in}
     \subfigure{\includegraphics[width=.40\textwidth]{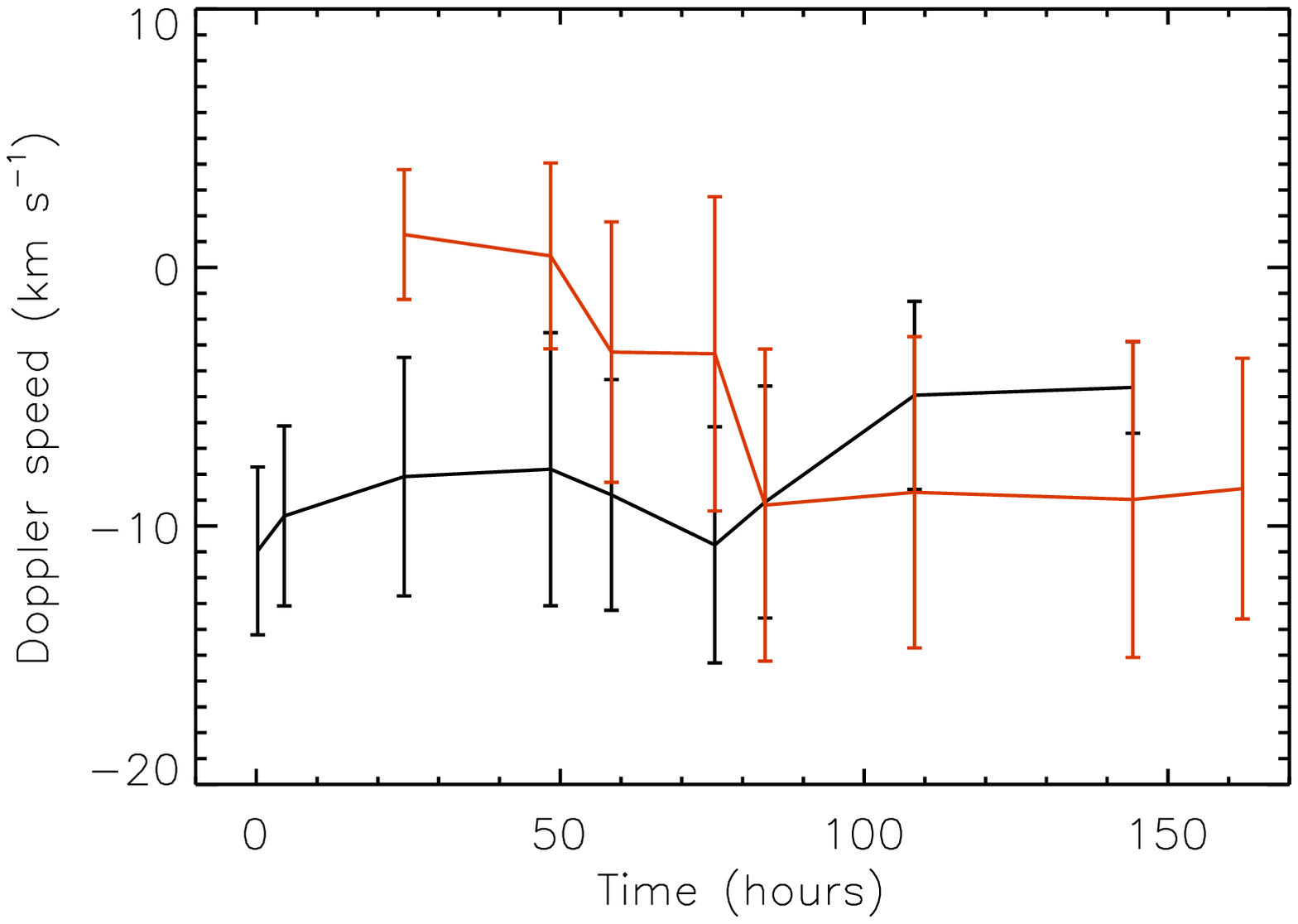}}
     \subfigure{\includegraphics[width=.40\textwidth]{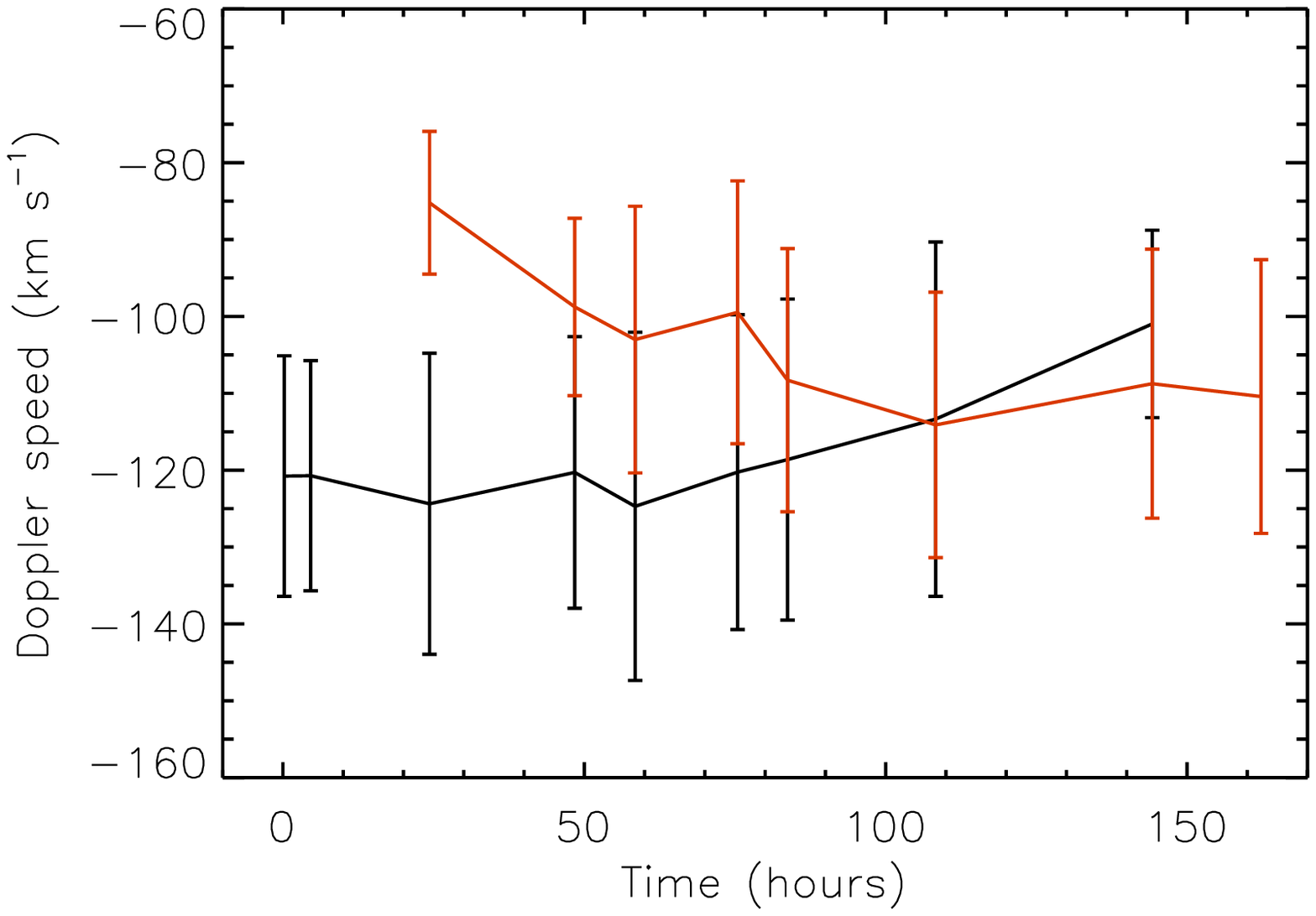}}
     \caption{Plots of average velocity with standard deviation versus time 
     over the duration of the EIS observations.
     Black lines represents average velocities from the western outflow
     region and red lines indicate those from the eastern outflow region.
     The left panel represents the primary Gaussian and the right represents
     the secondary Gaussian.
     Time is given in hours from 2007 Dec 09 00:00:00 UT.}
     \label{fig:variation}
\end{figure}

Outflows are found to persist for the 6 day duration of the observations. In
Figure~\ref{fig:variation}, we show the variations in the derived flow velocities
over this period.  Here, we have compared the velocities of the primary and
secondary flow components in both the eastern and western outflowing regions, and
also indicated the standard deviation of velocities throughout the respective
regions. Despite the observations spanning a significant extent of the solar
rotation, there is relatively little change in the line-of-sight velocities over
this time.  The western outflow region (black lines in
Figure~\ref{fig:variation}) show a slight decreasing trend with time (for both
the primary and secondary components). The eastern outflow region (red lines)
shows a more pronounced, increasing velocity, trend over time.   From the
geometry of the AR, these apparent flow speeds are what we would  expect,
although we do note that there is little statistical significance to the trends.
It seems evident that the outflow is not confined to a narrow cross-sectional
area but, rather, emanates over a wide `cone' of emission. The large standard
deviation in the velocities shown in Figure~\ref{fig:variation} further supports
this interpretation. As discussed in Section~\ref{sec:results}, the outflowing
regions are also obscured by the overlying AR loop system, which further
complicates the analysis.

\begin{figure}[htp]
     \centering
     \vspace{-.45in}
     \subfigure{\includegraphics[width=.40\textwidth]{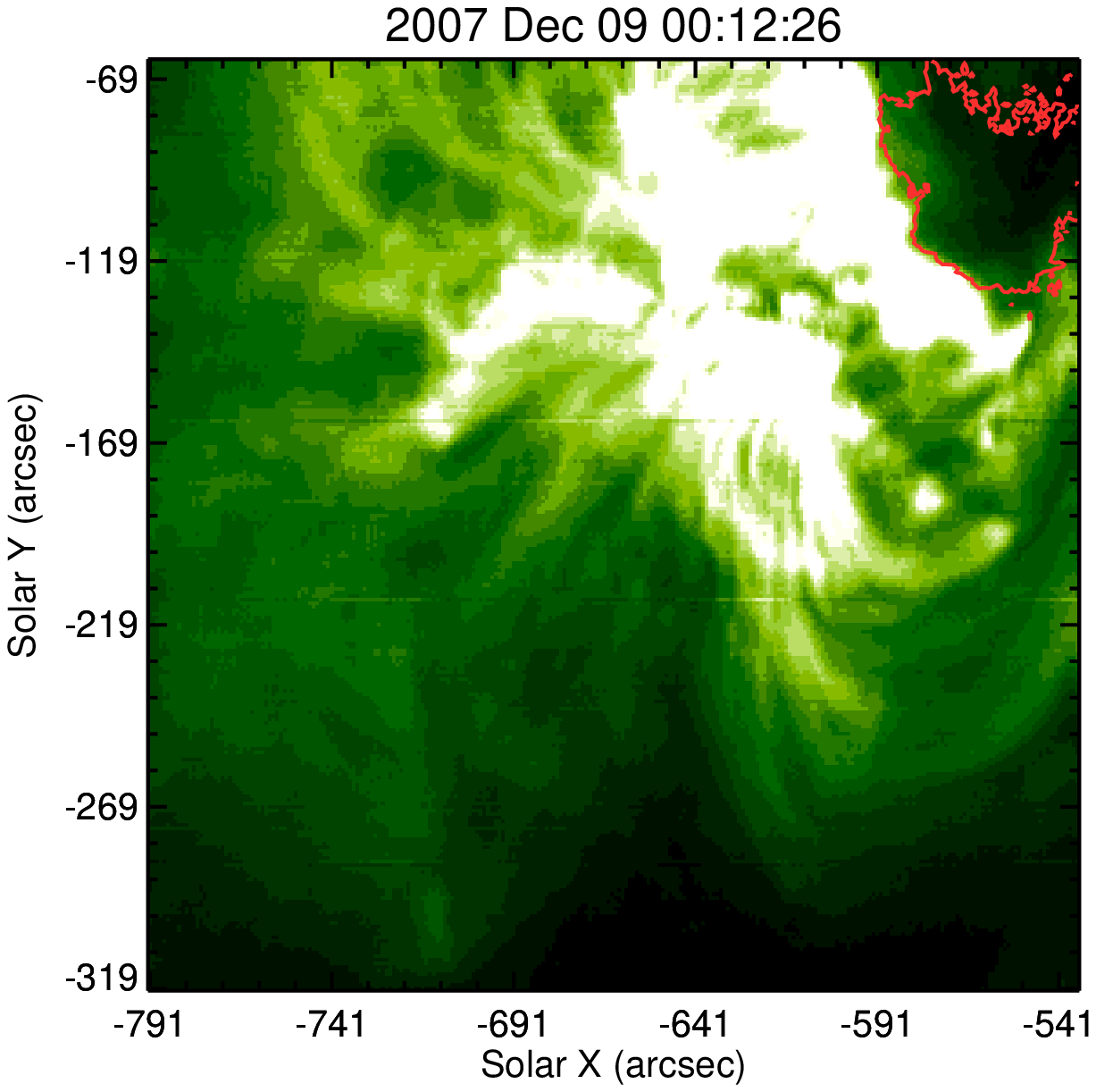}}
     \subfigure{\includegraphics[width=.40\textwidth]{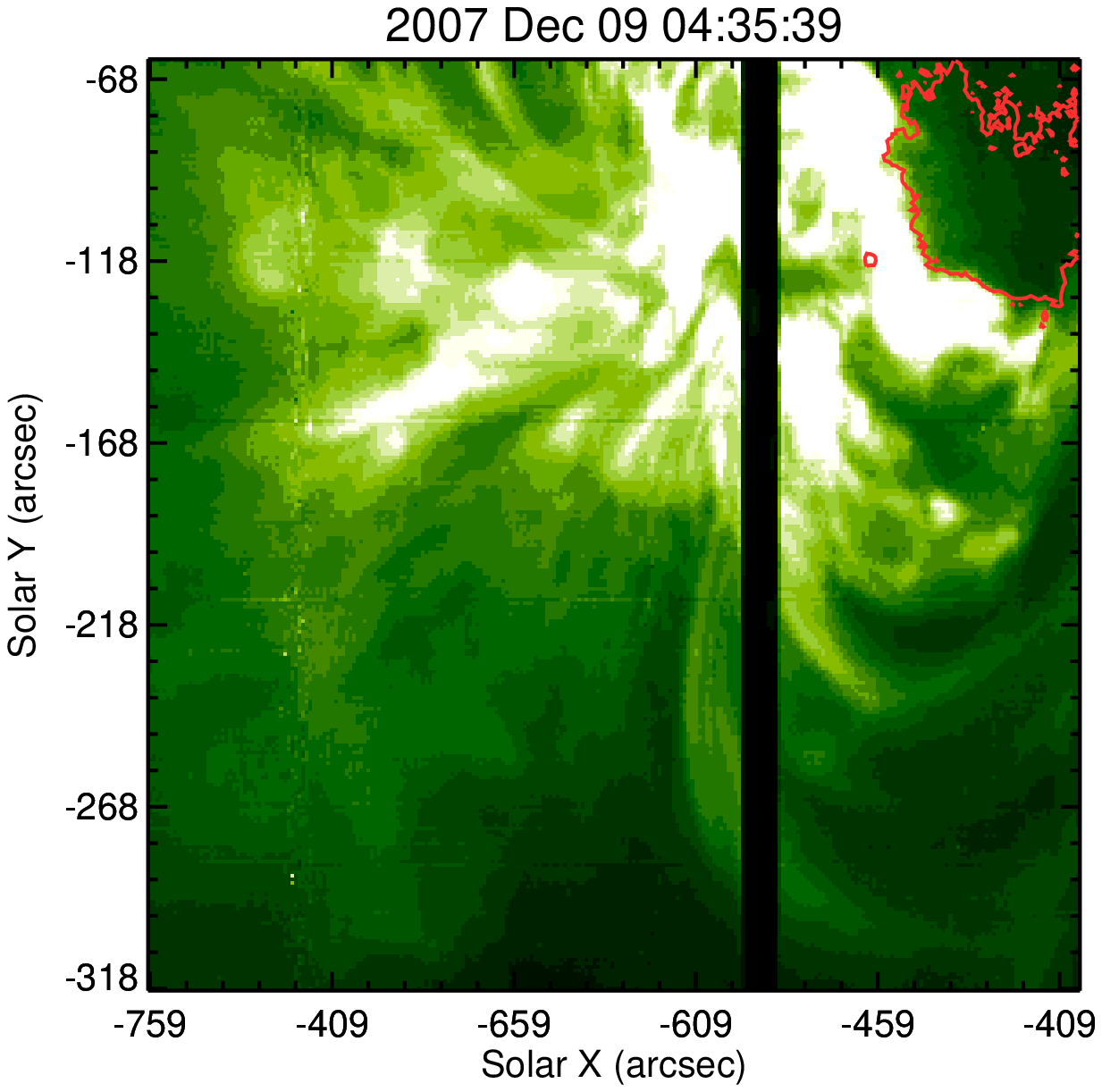}}\\
     \vspace{-.2in}
     \subfigure{\includegraphics[width=.40\textwidth]{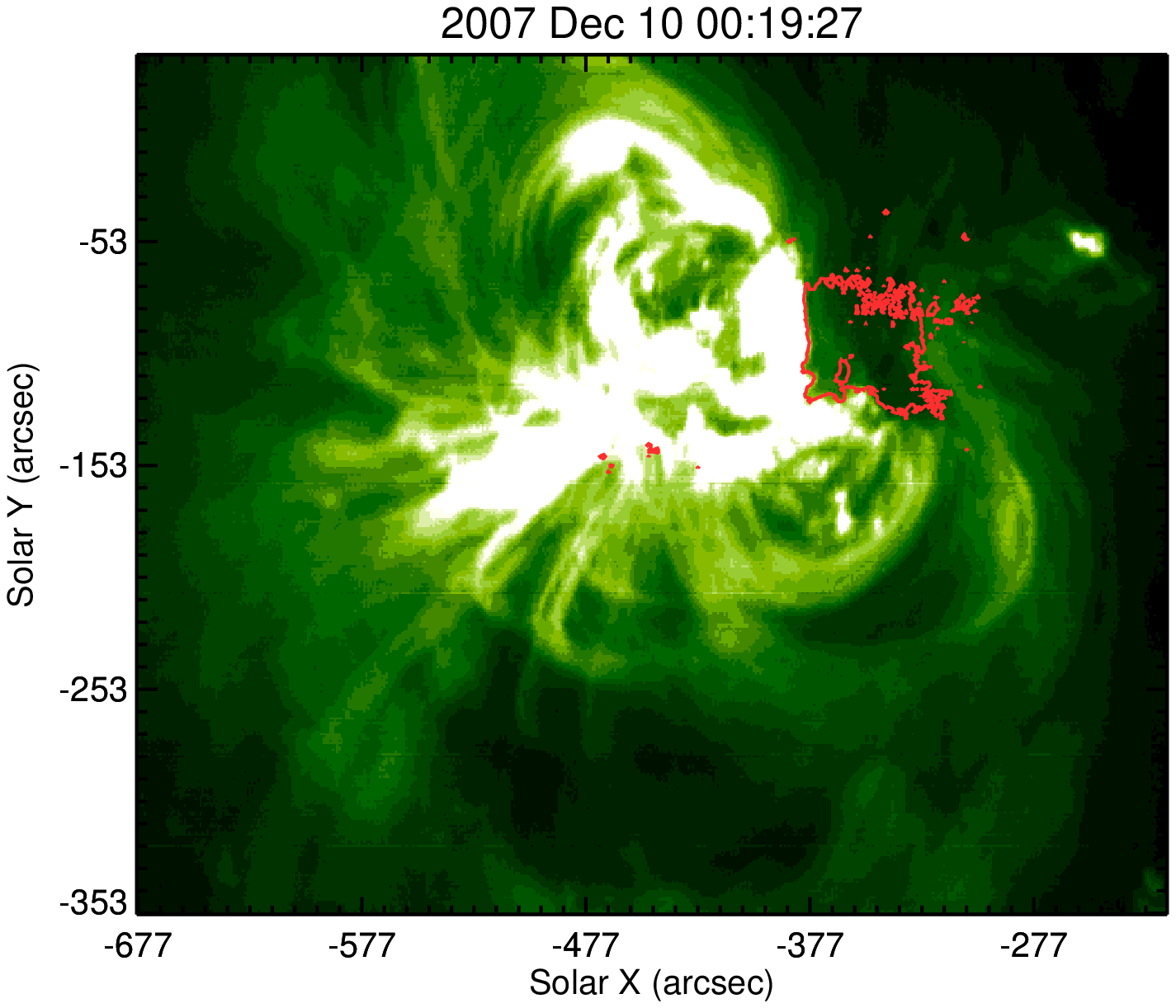}}
     \subfigure{\includegraphics[width=.40\textwidth]{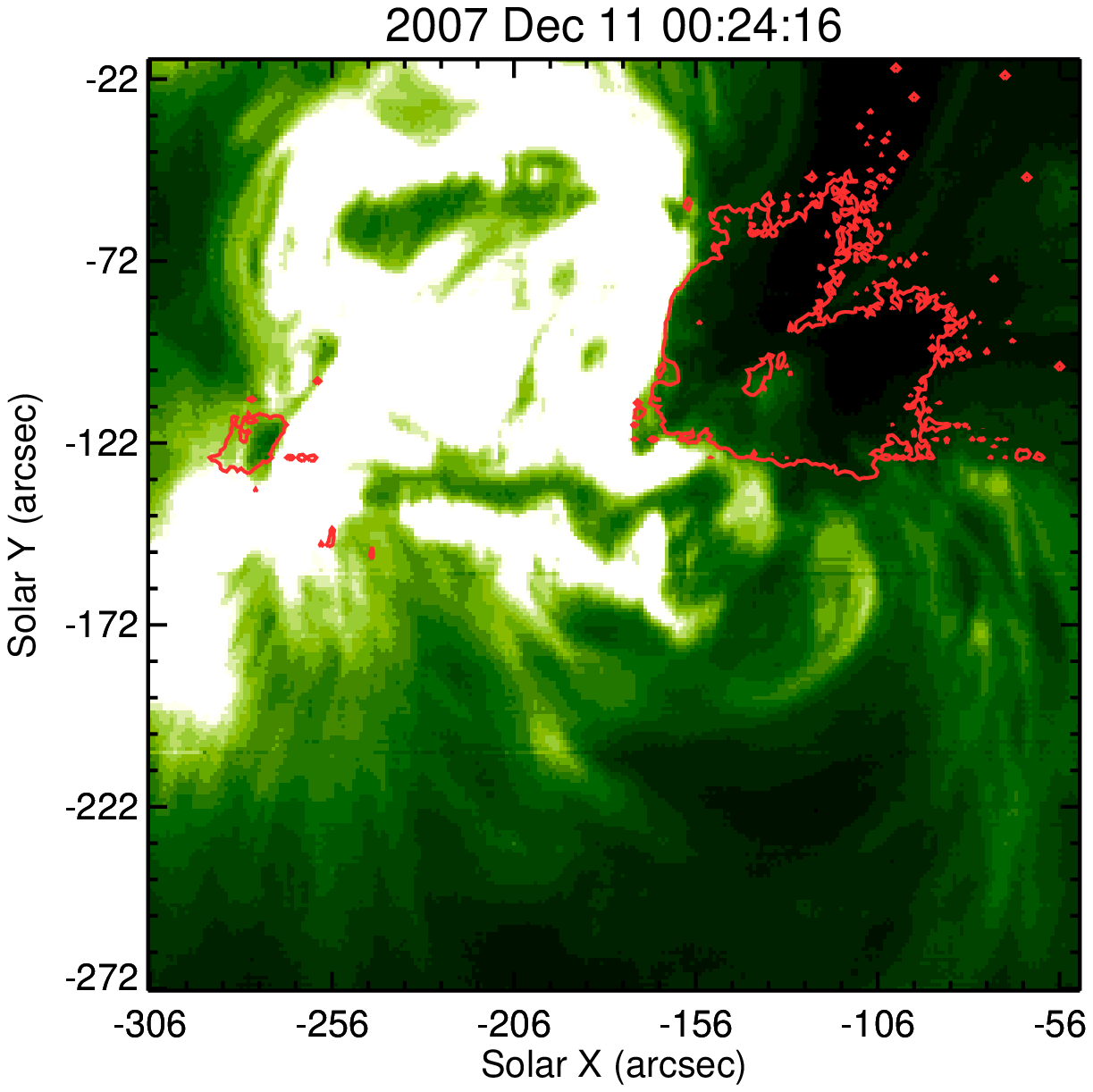}}\\
     \vspace{-.2in}
     \subfigure{\includegraphics[width=.40\textwidth]{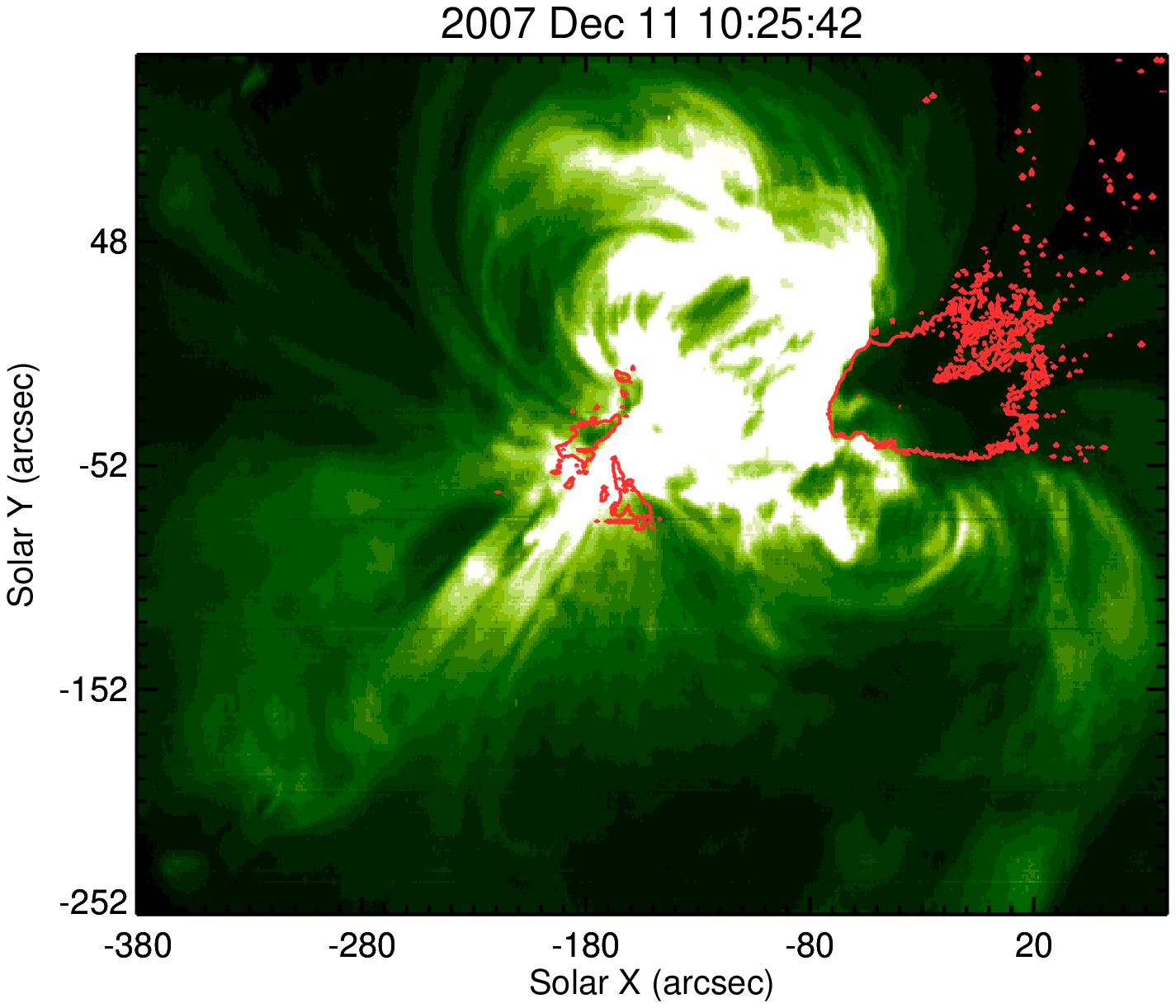}}
     \subfigure{\includegraphics[width=.40\textwidth]{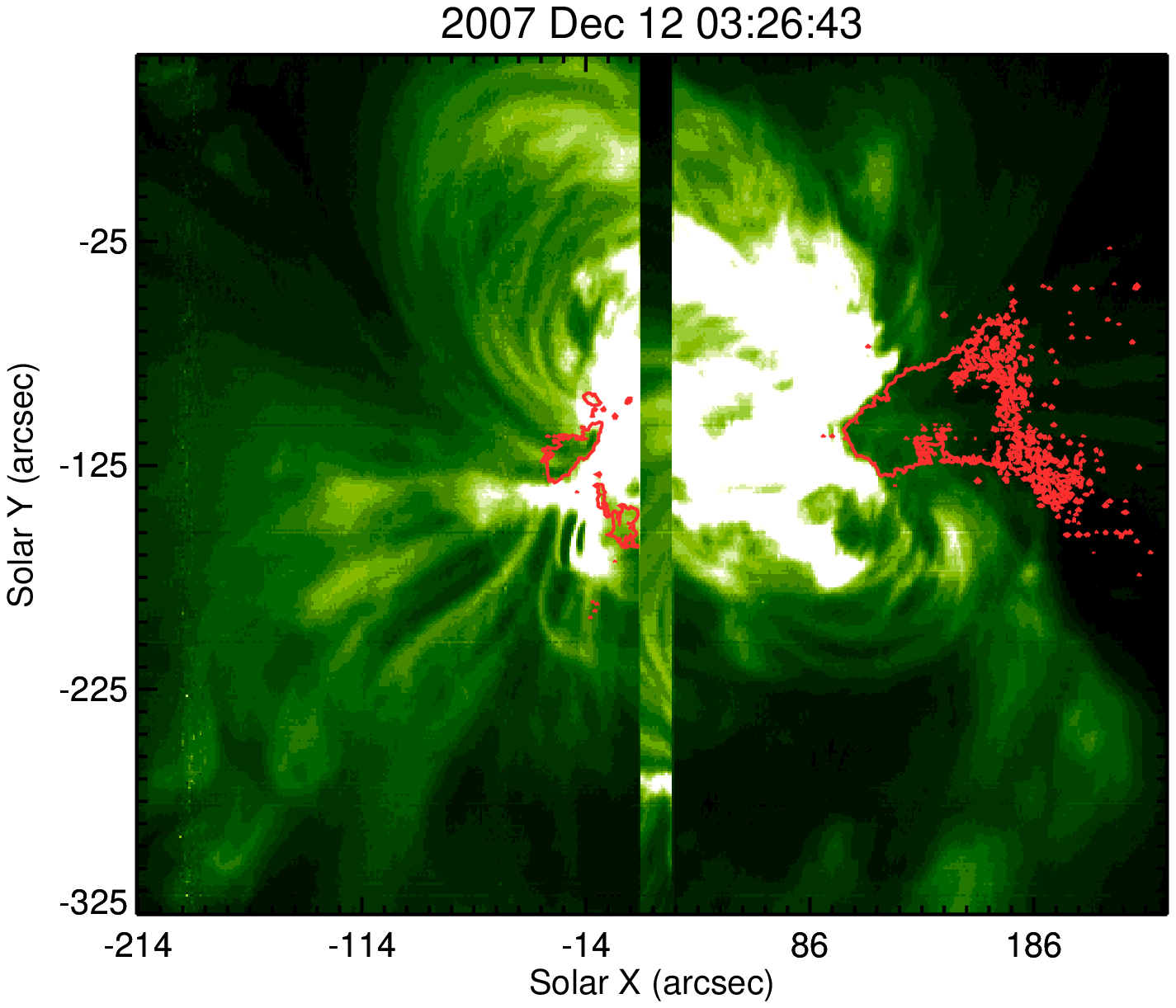}}\\
     \vspace{-.2in}
     \subfigure{\includegraphics[width=.40\textwidth]{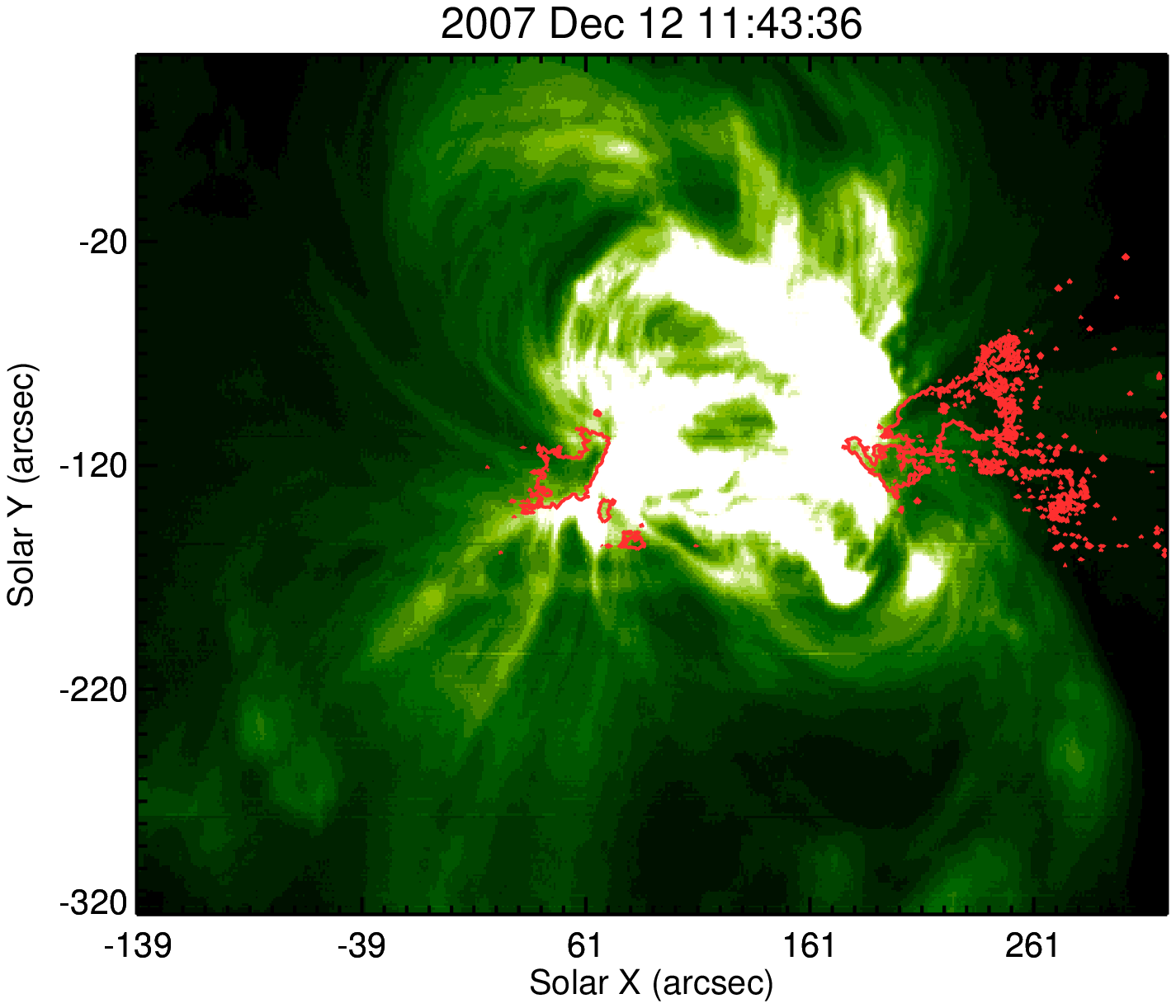}}
     \subfigure{\includegraphics[width=.40\textwidth]{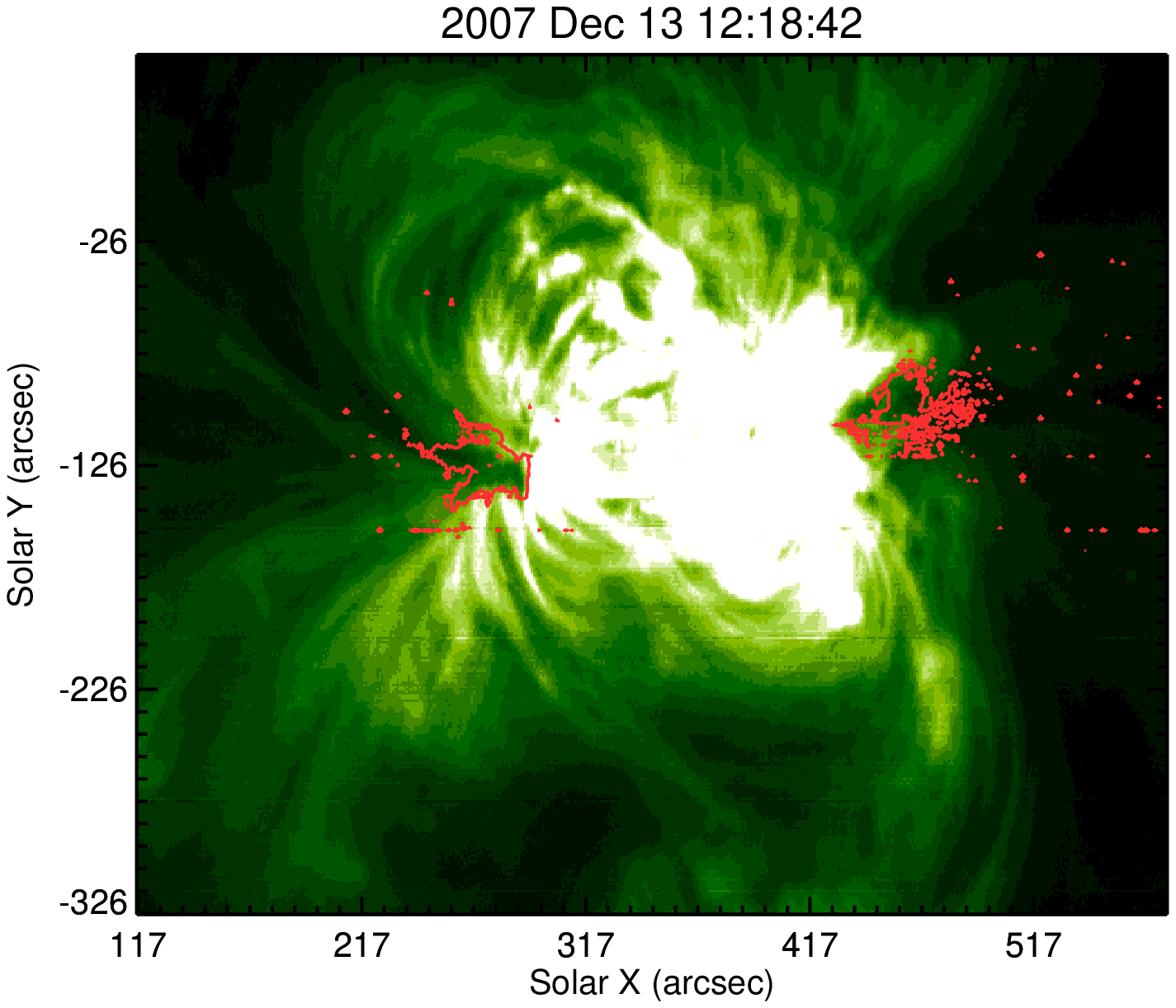}}\\
     \vspace{-.2in}
     \subfigure{\includegraphics[width=.40\textwidth]{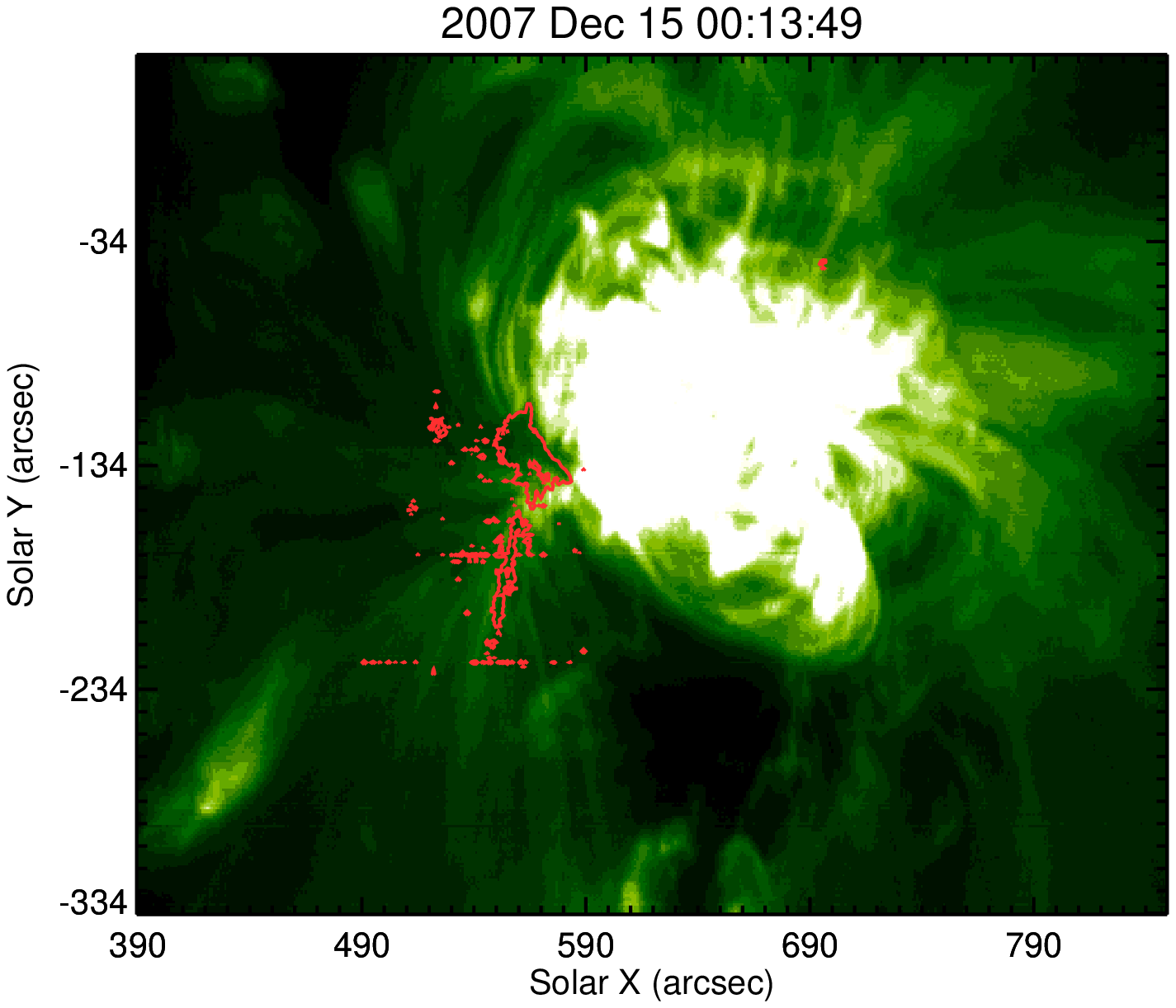}}
     \subfigure{\includegraphics[width=.40\textwidth]{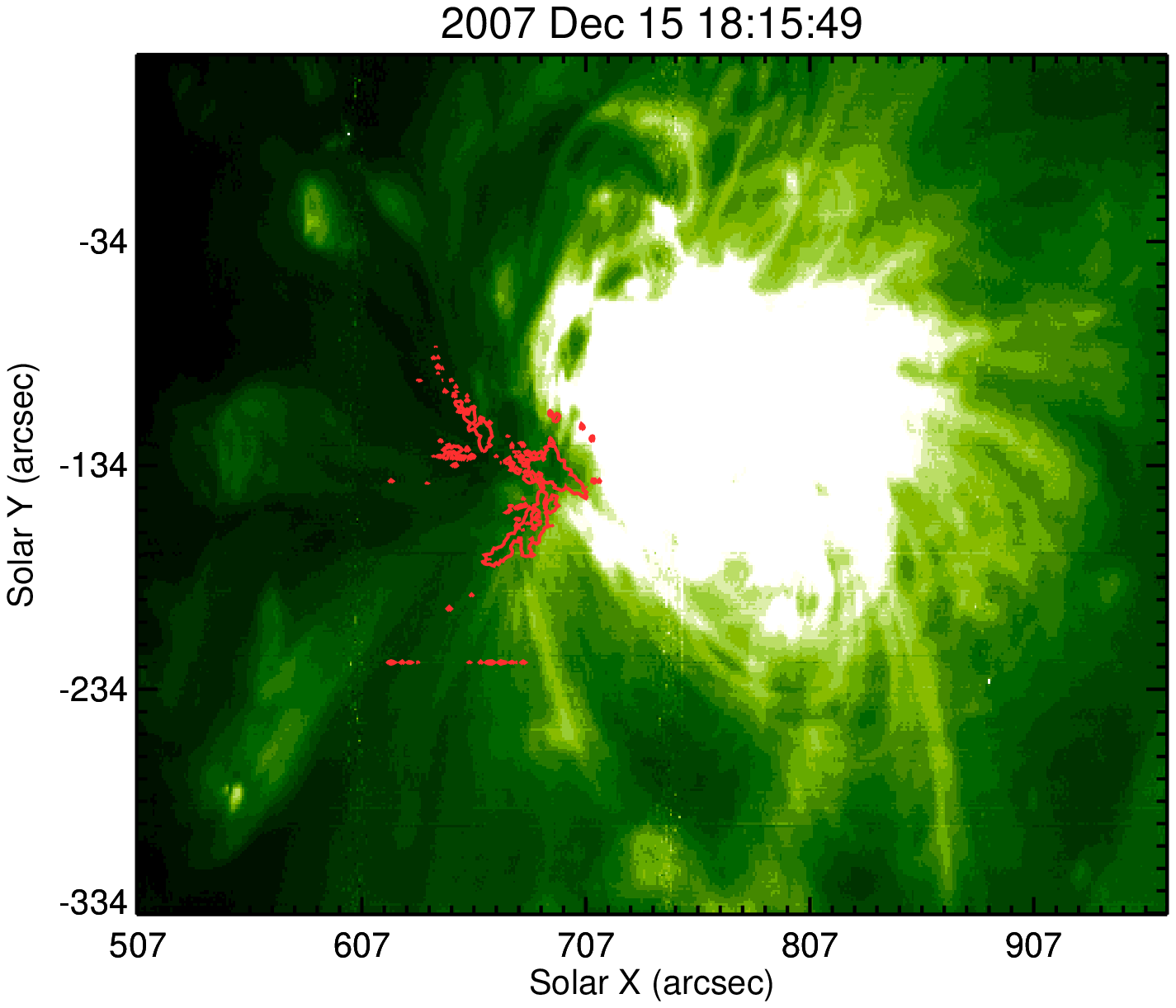}}
     \caption{
     Intensity images for each of the 10 EIS observations.
     Time runs from left to right and top to bottom.  The intensities are 
     the sum of the intensities of the two Gaussians.
     The red contours show where the intensity of the secondary component 
     is 5\% of the primary.
     The abscissa and ordinates are in units of arcsec from Sun center.}
     \label{fig:intensity}
\end{figure}

\begin{figure}[htp]
     \centering
     \vspace{-.45in}
     \subfigure{\includegraphics[width=.40\textwidth]{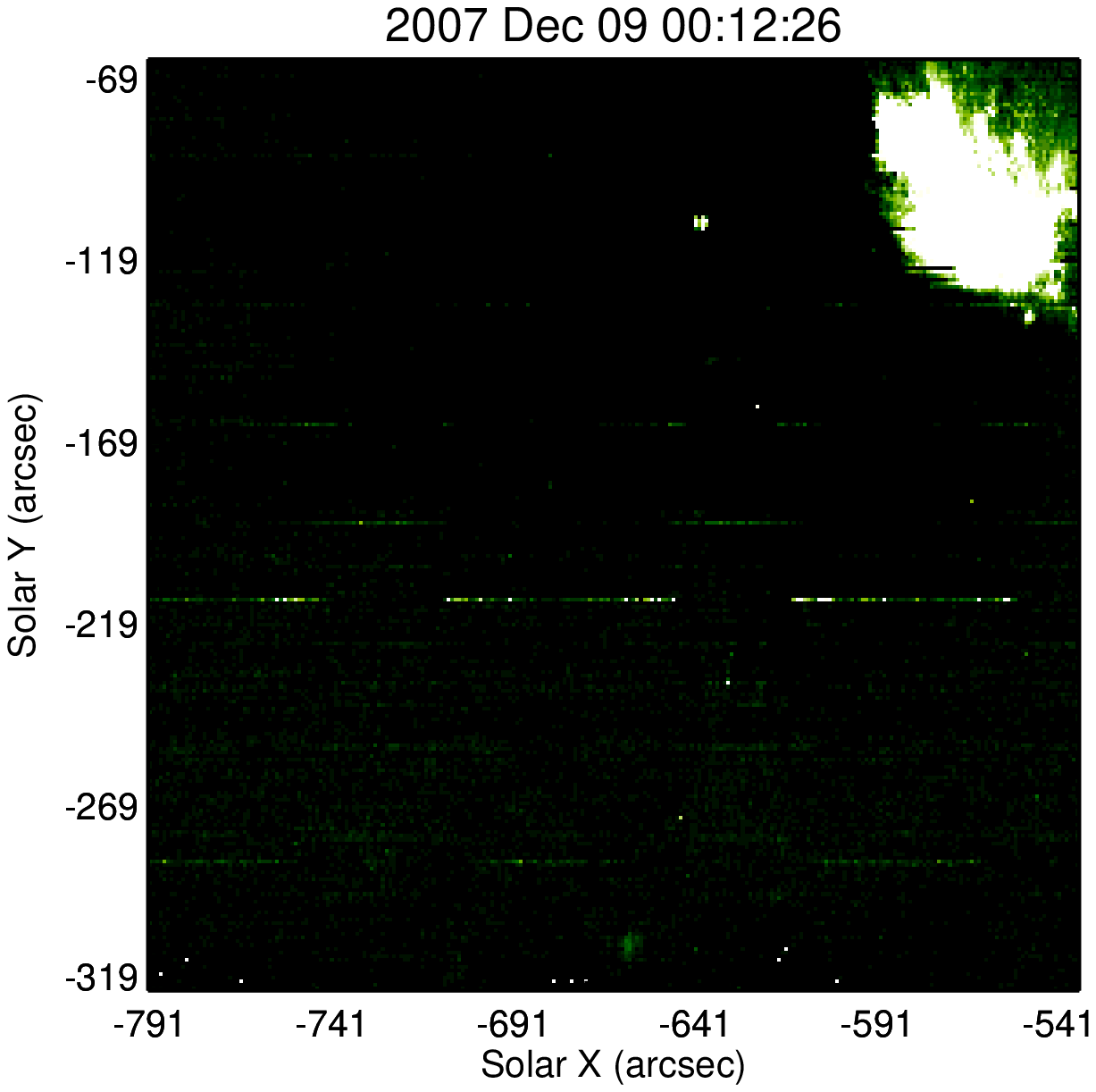}}
     \subfigure{\includegraphics[width=.40\textwidth]{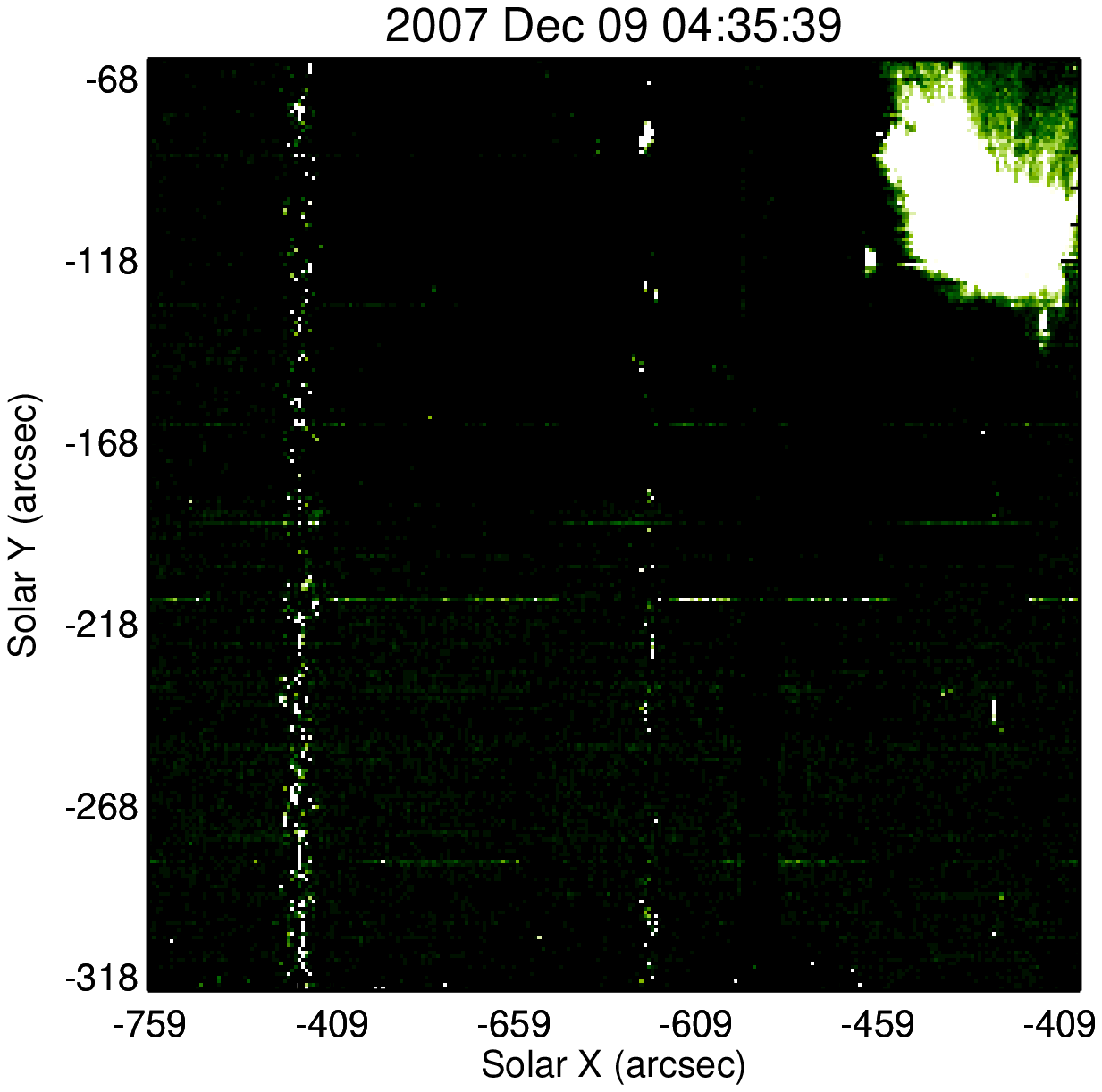}}\\
     \vspace{-.2in}
     \subfigure{\includegraphics[width=.40\textwidth]{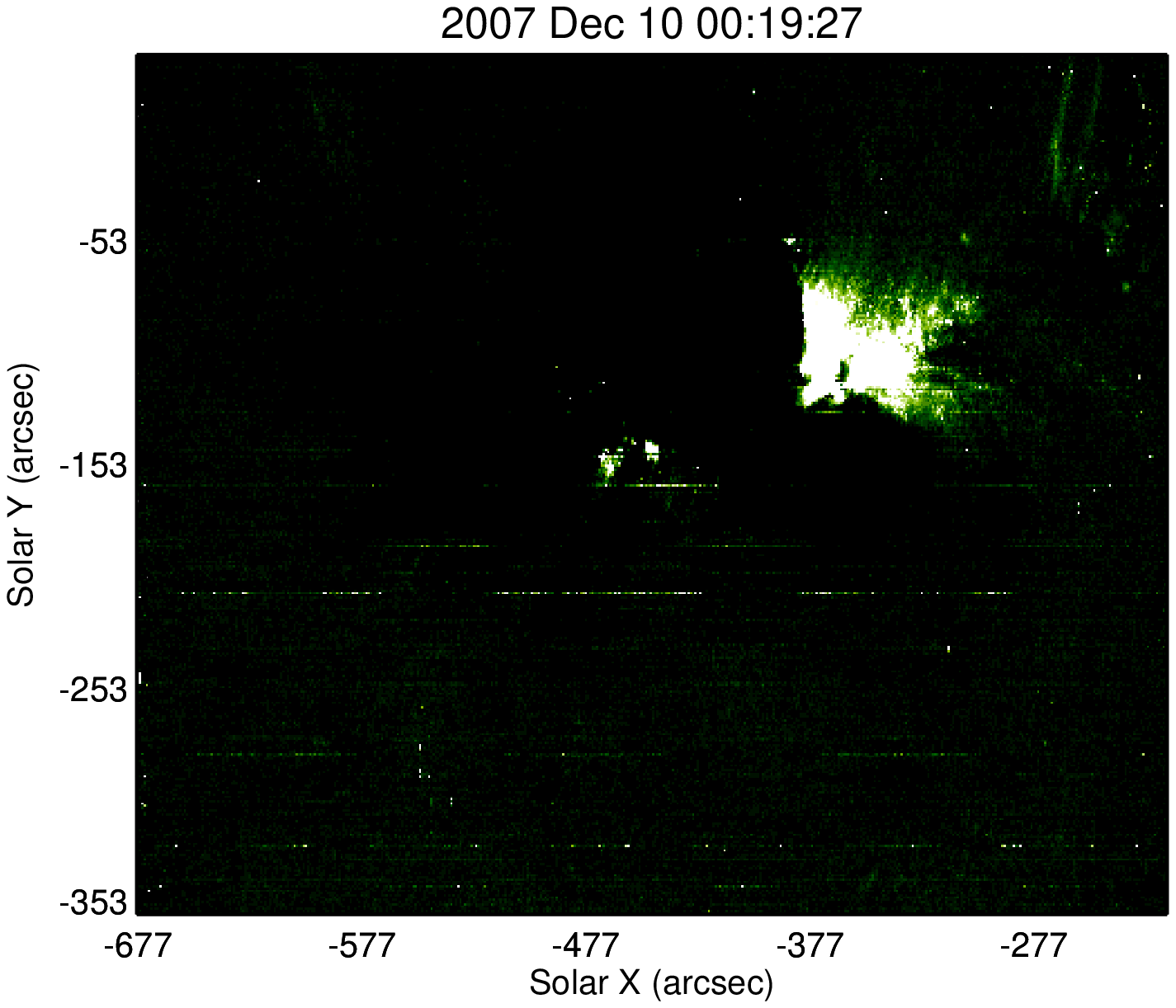}}
     \subfigure{\includegraphics[width=.40\textwidth]{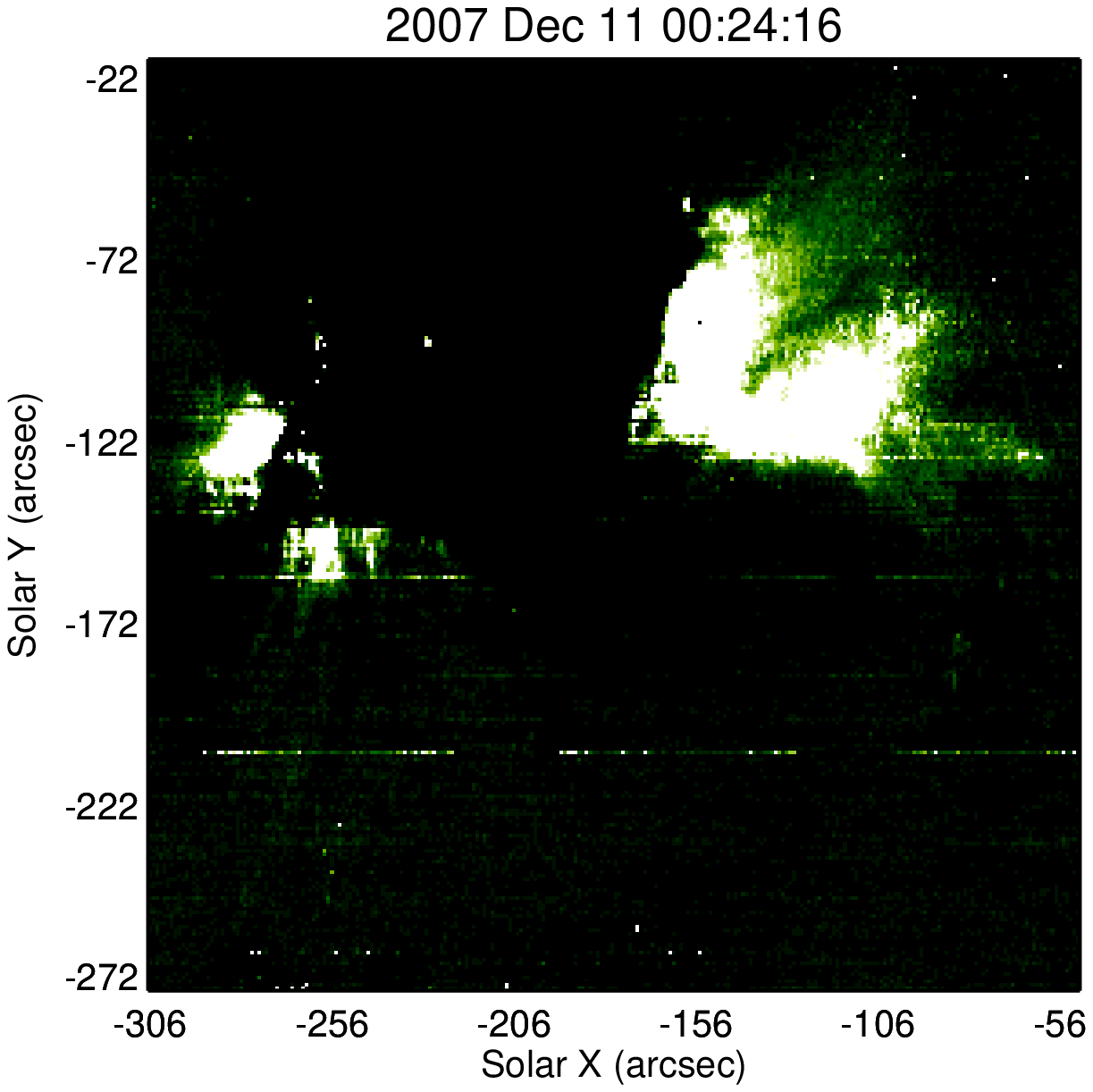}}\\
     \vspace{-.2in}
     \subfigure{\includegraphics[width=.40\textwidth]{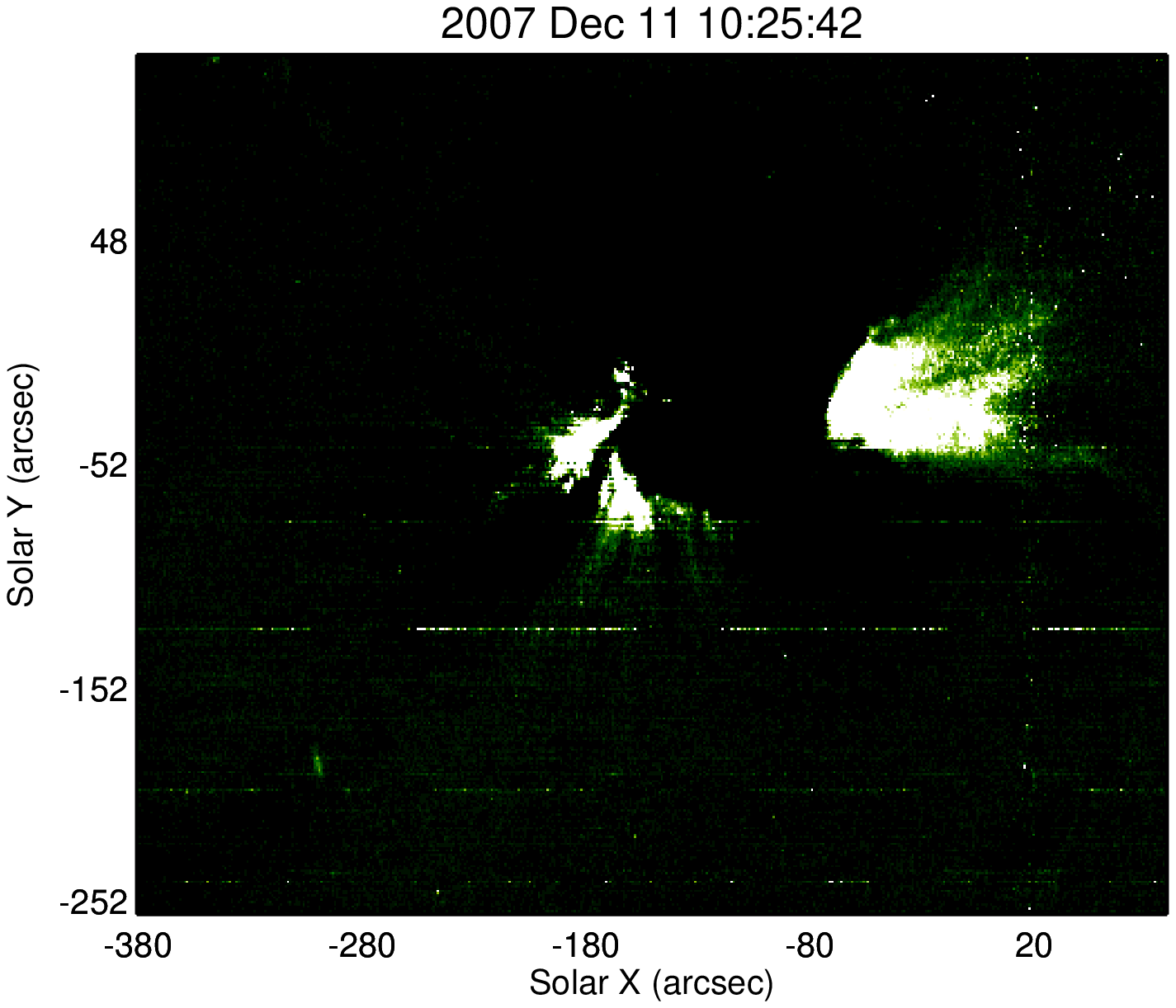}}
     \subfigure{\includegraphics[width=.40\textwidth]{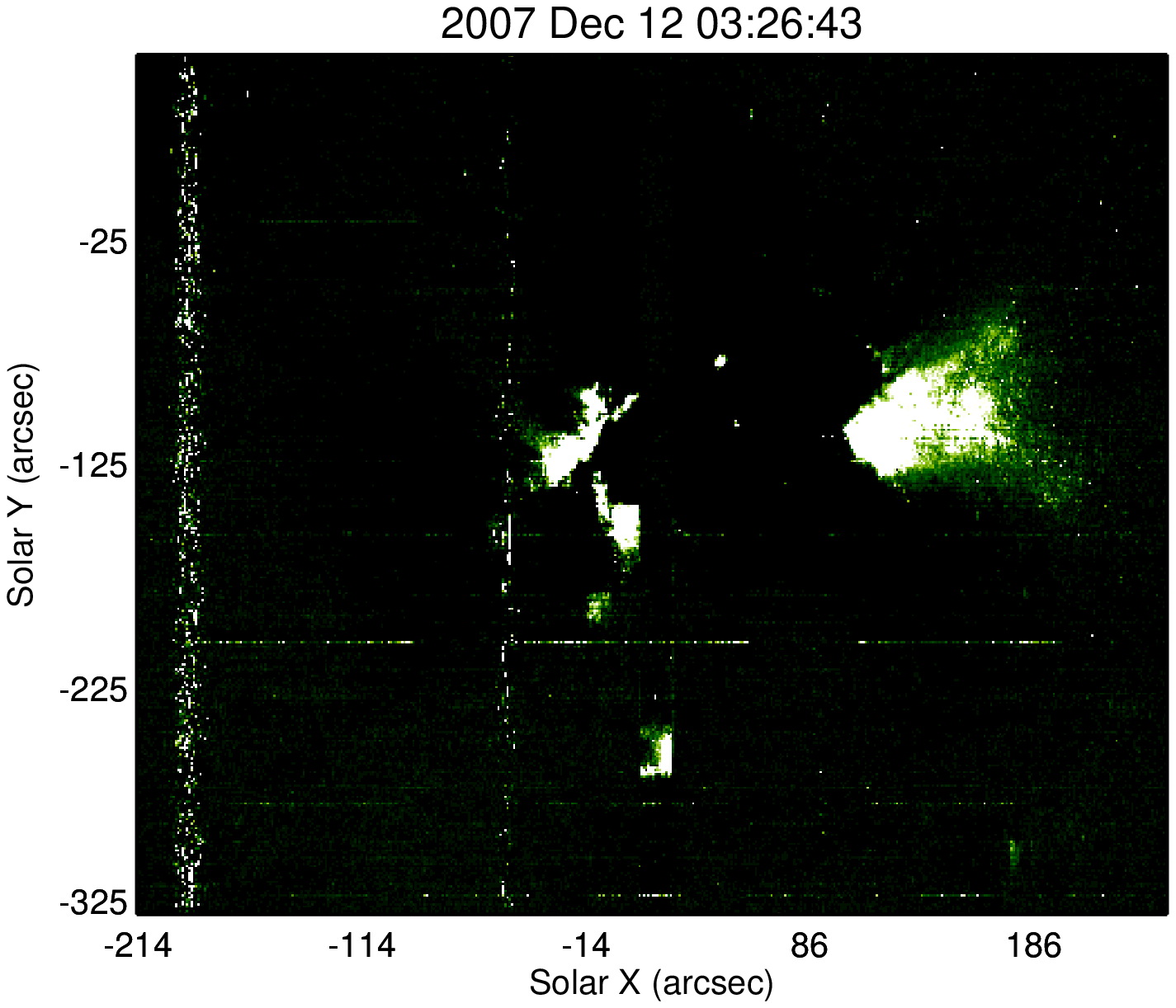}}\\
     \vspace{-.2in}
     \subfigure{\includegraphics[width=.40\textwidth]{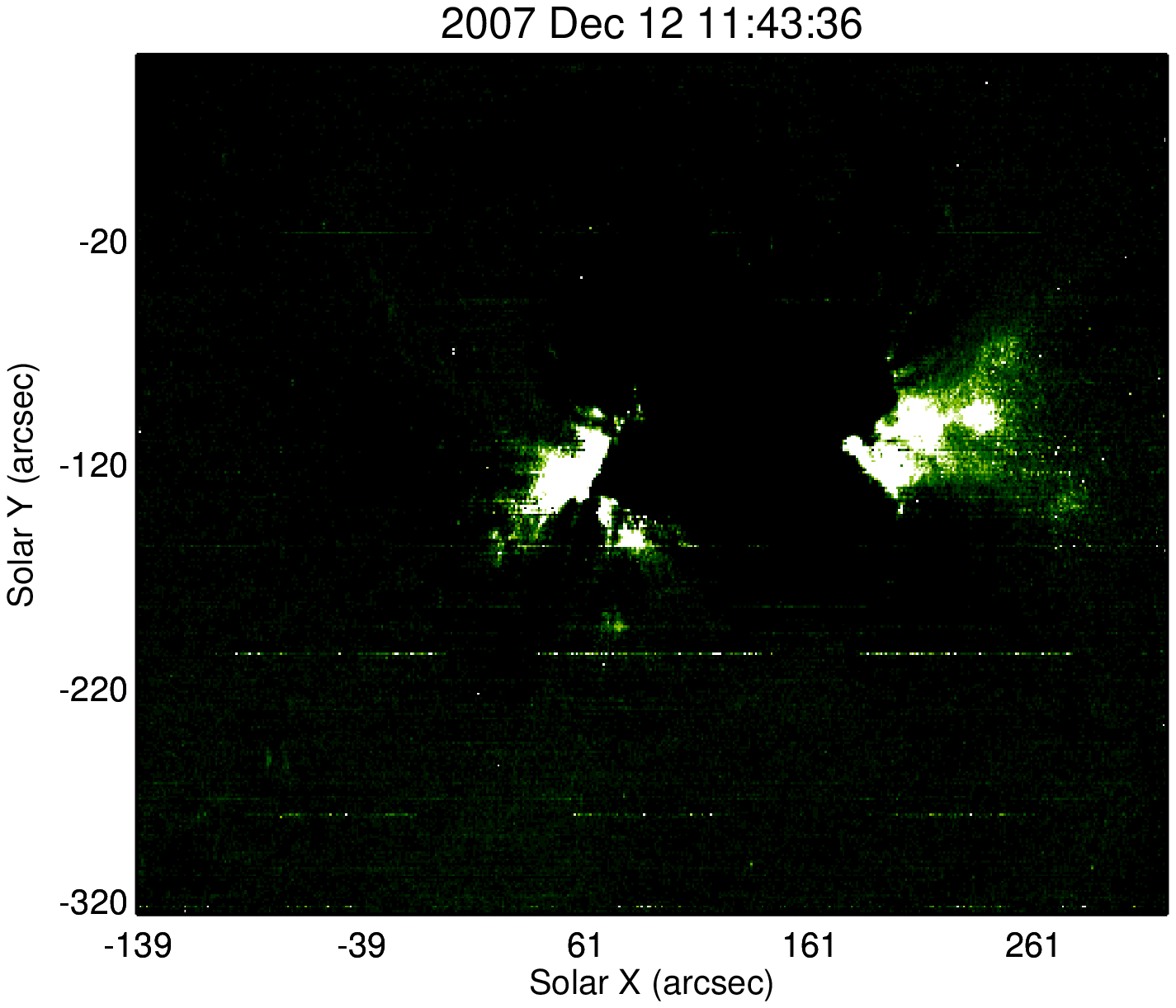}}
     \subfigure{\includegraphics[width=.40\textwidth]{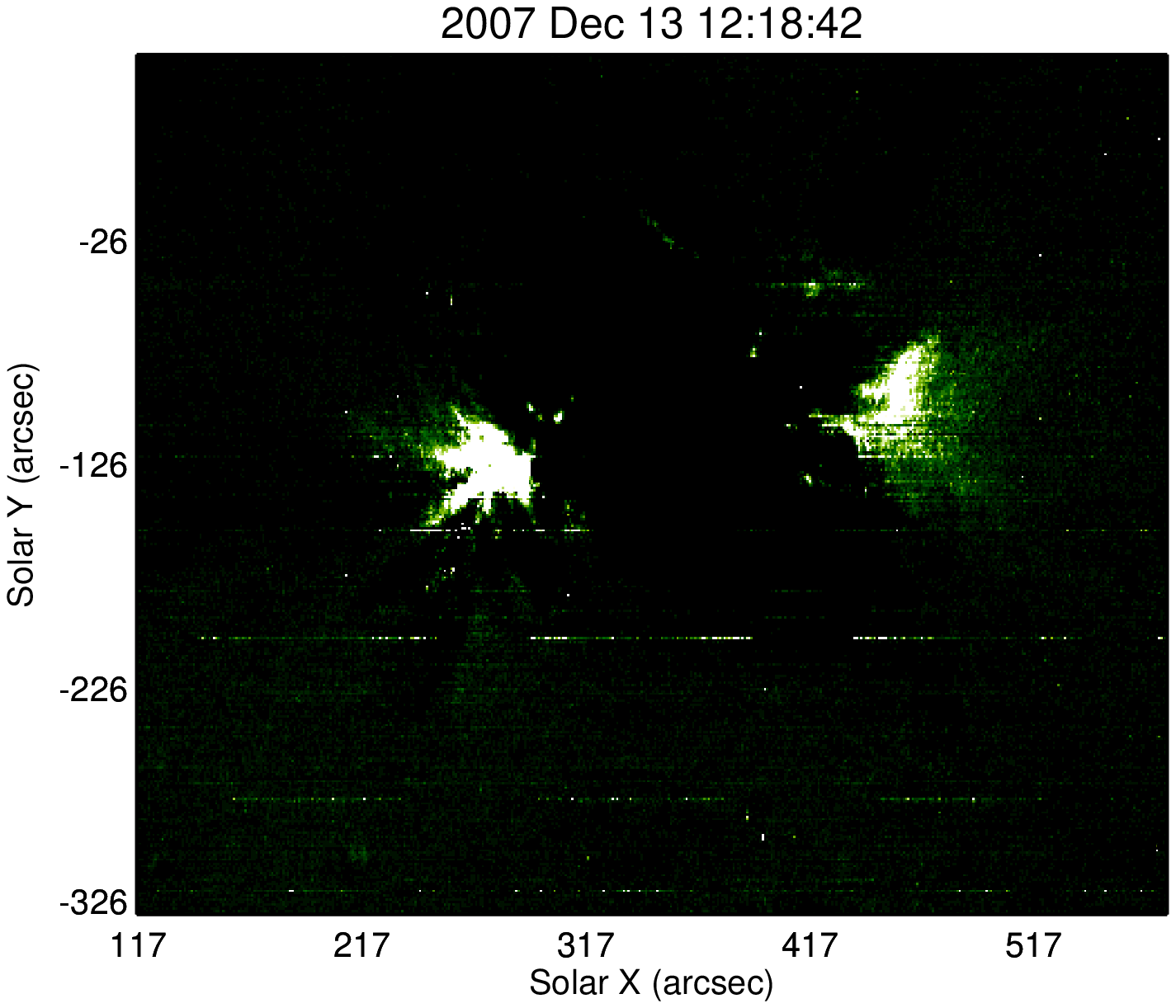}}\\
     \vspace{-.2in}
     \subfigure{\includegraphics[width=.40\textwidth]{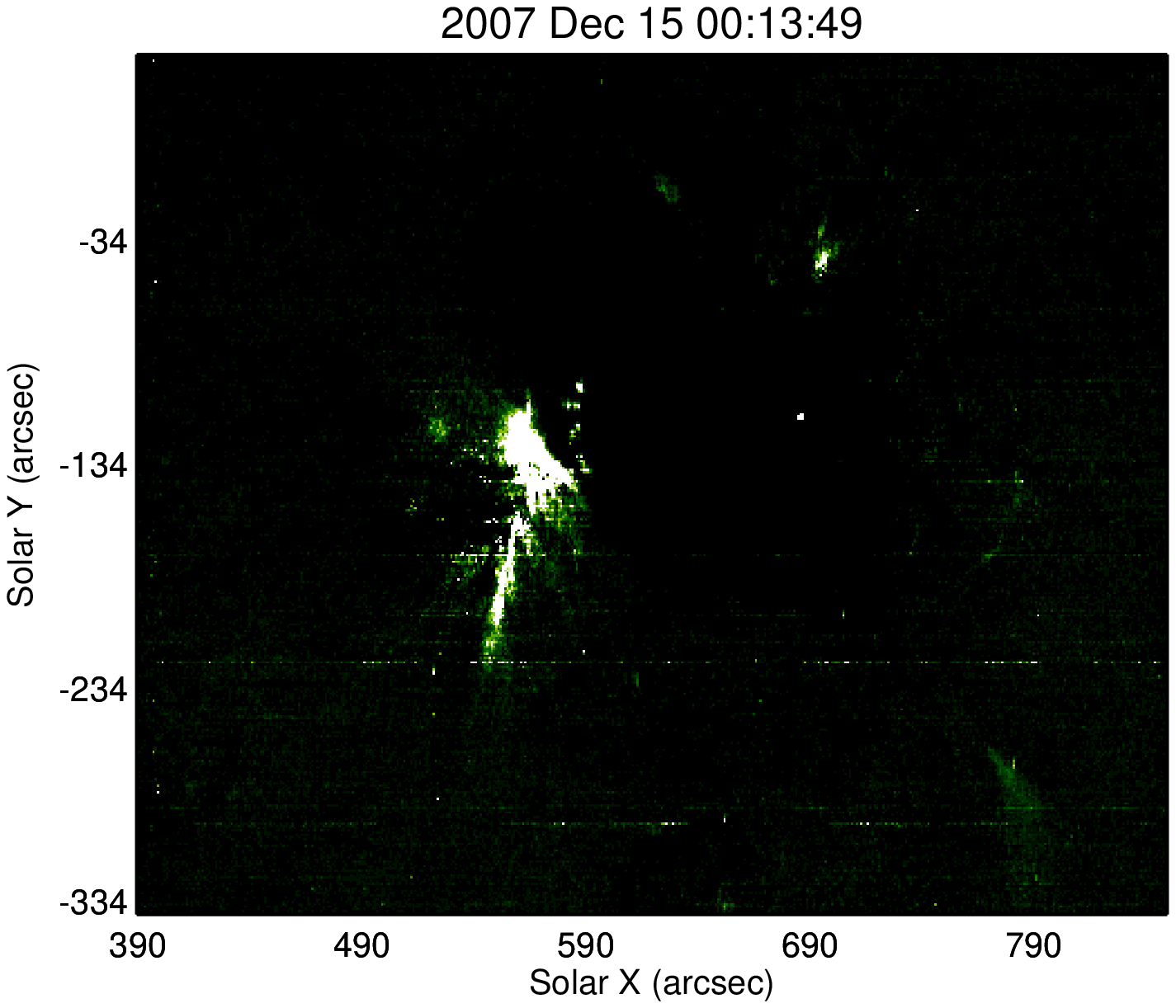}}
     \subfigure{\includegraphics[width=.40\textwidth]{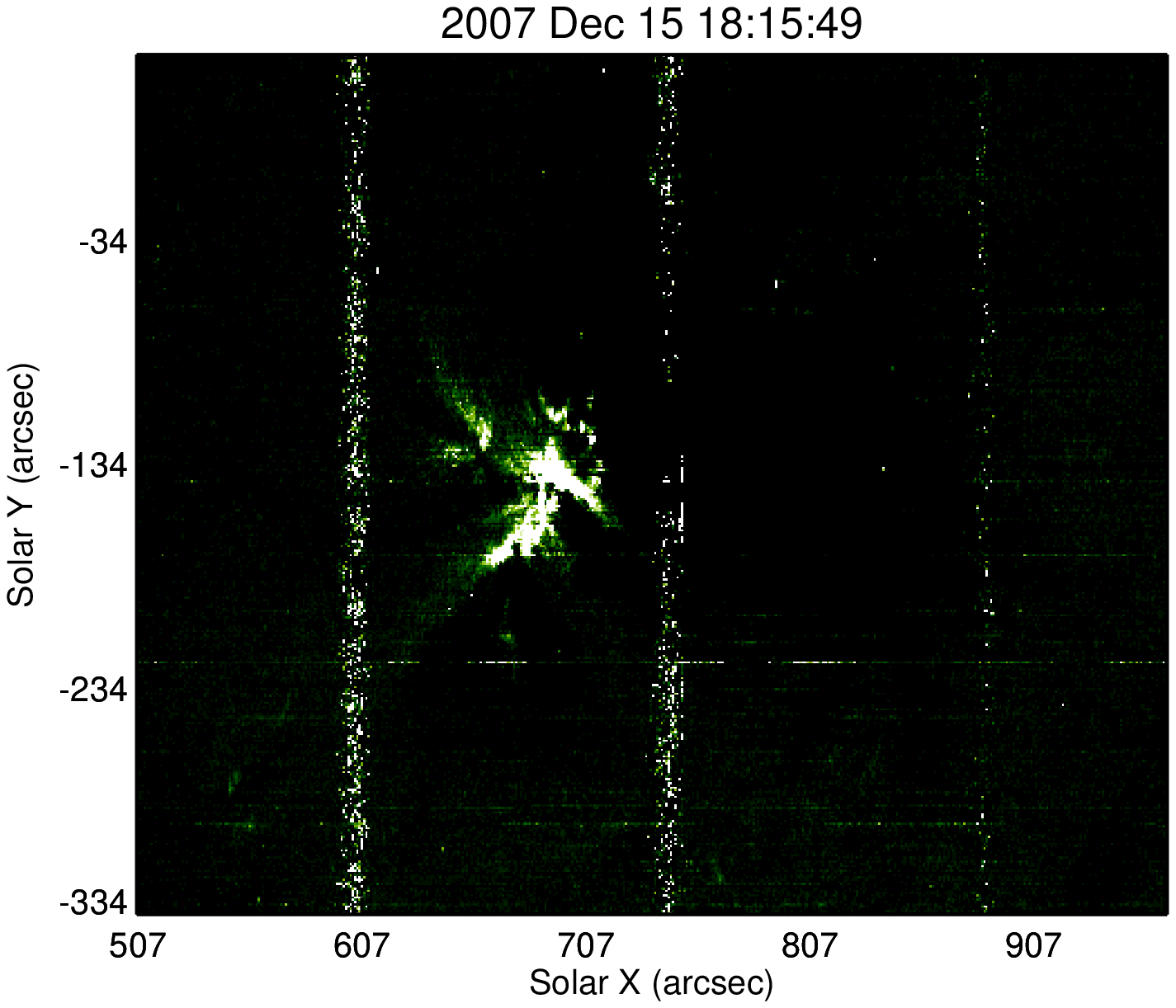}}
     \caption{
     Same as Figure~\protect\ref{fig:intensity} but only showing
     the intensity of the secondary Gaussian.}
     \label{fig:intensity secondary}
\end{figure}

\begin{figure}[htp]
     \centering
     \vspace{-1in}
     \includegraphics[width=.80\textwidth]{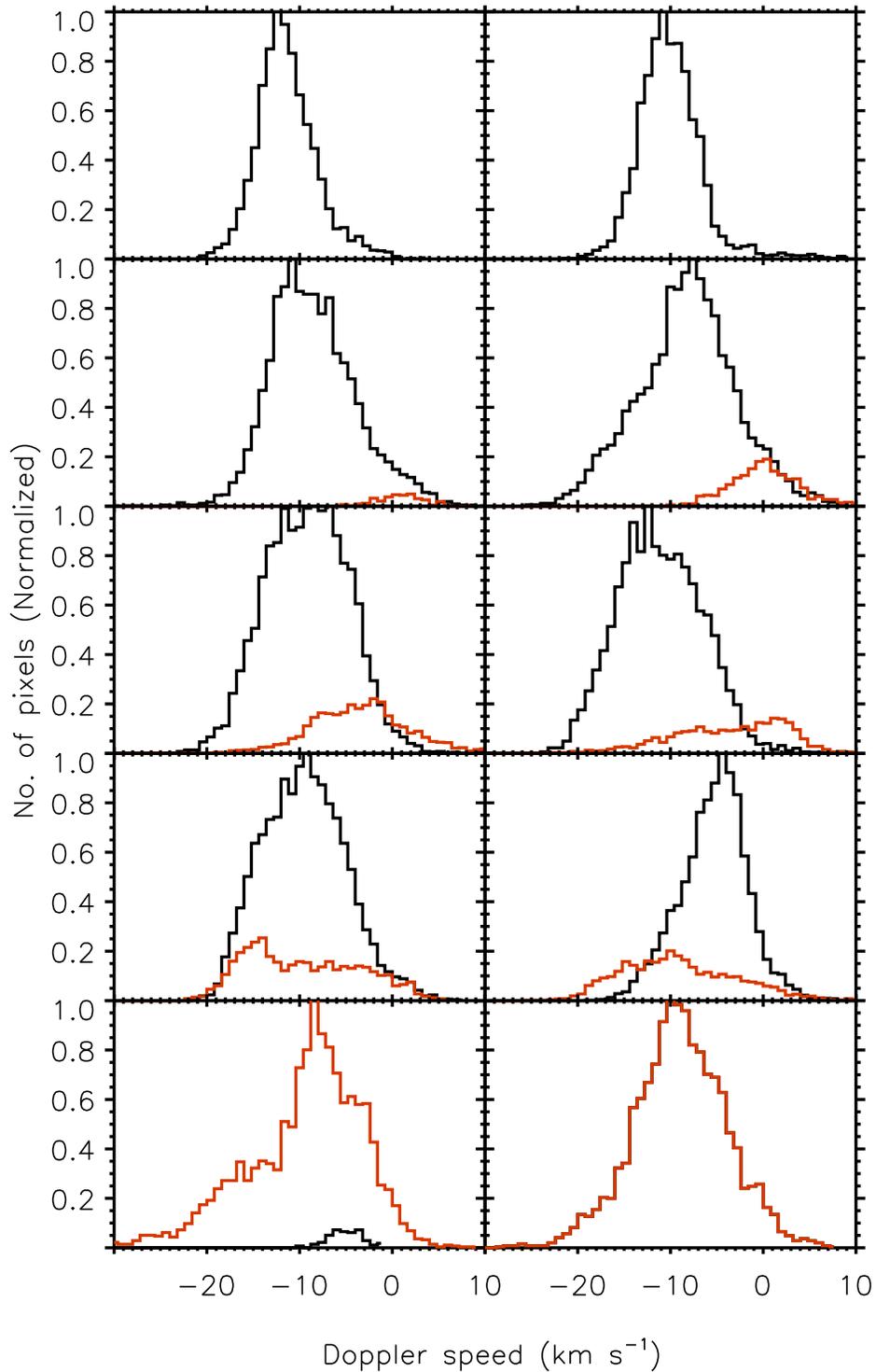}
     \caption{Histograms of the velocities of the primary Gaussian for
     each of the 10 EIS observations.  Black lines represent the western
     outflow region and red lines represent the eastern outflow region.
     Only pixels where the intensity of the secondary component
     is at least 10\% of the intensity of the primary component are plotted.
     The plots follow the same temporal order as
     Figure~\protect\ref{fig:intensity}; left to right and top to bottom.}
     \label{fig:histogram main}
\end{figure}

\begin{figure}[htp]
     \centering
     \vspace{-1in}
     \includegraphics[width=.80\textwidth]{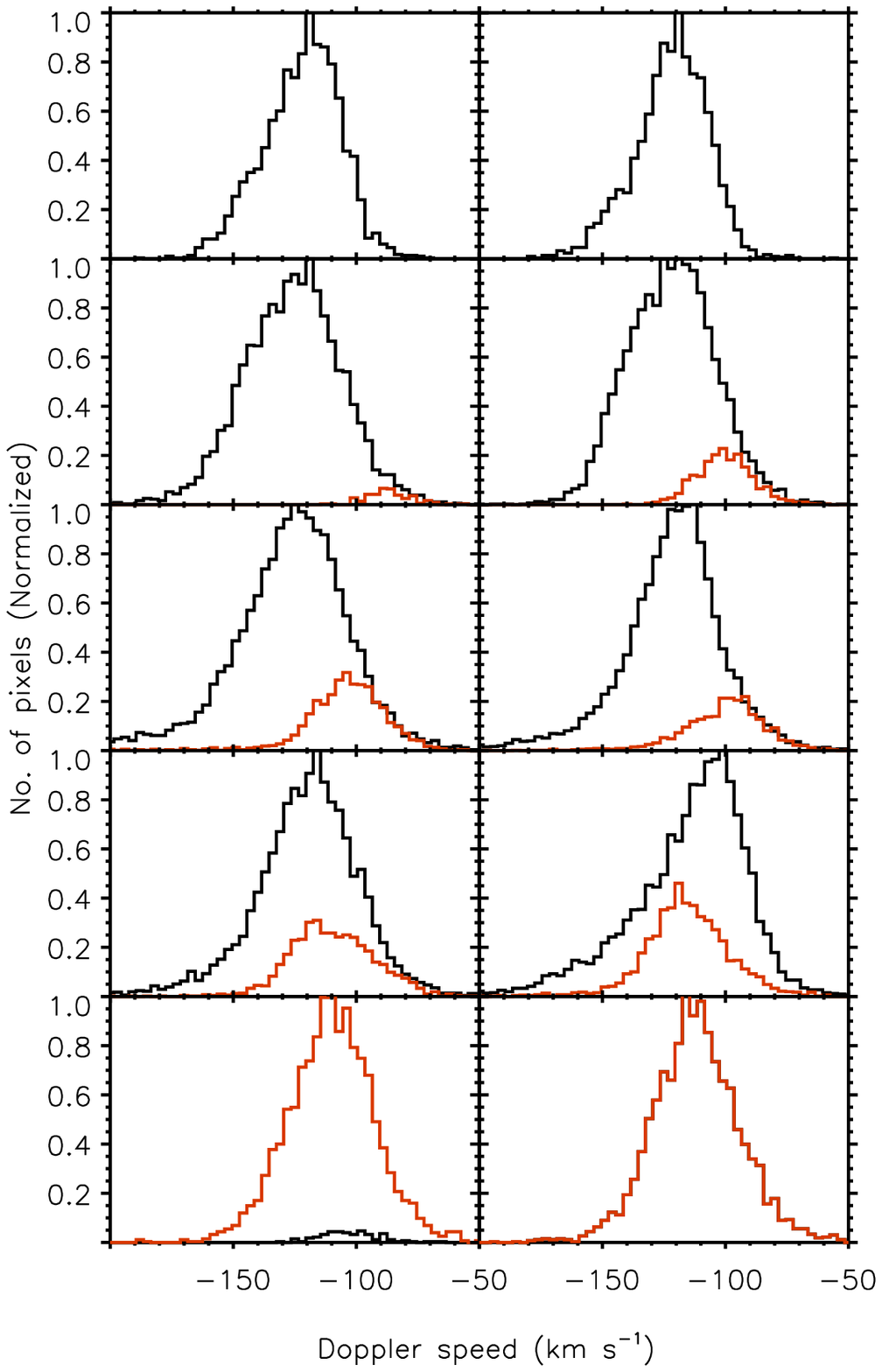}
     \caption{Same as Figure~\protect\ref{fig:histogram main} but for the
     secondary Gaussian component.}
     \label{fig:histogram minor}
\end{figure}

\begin{figure}[htp]
     \centering
     \vspace{-1in}
     \includegraphics[width=.80\textwidth]{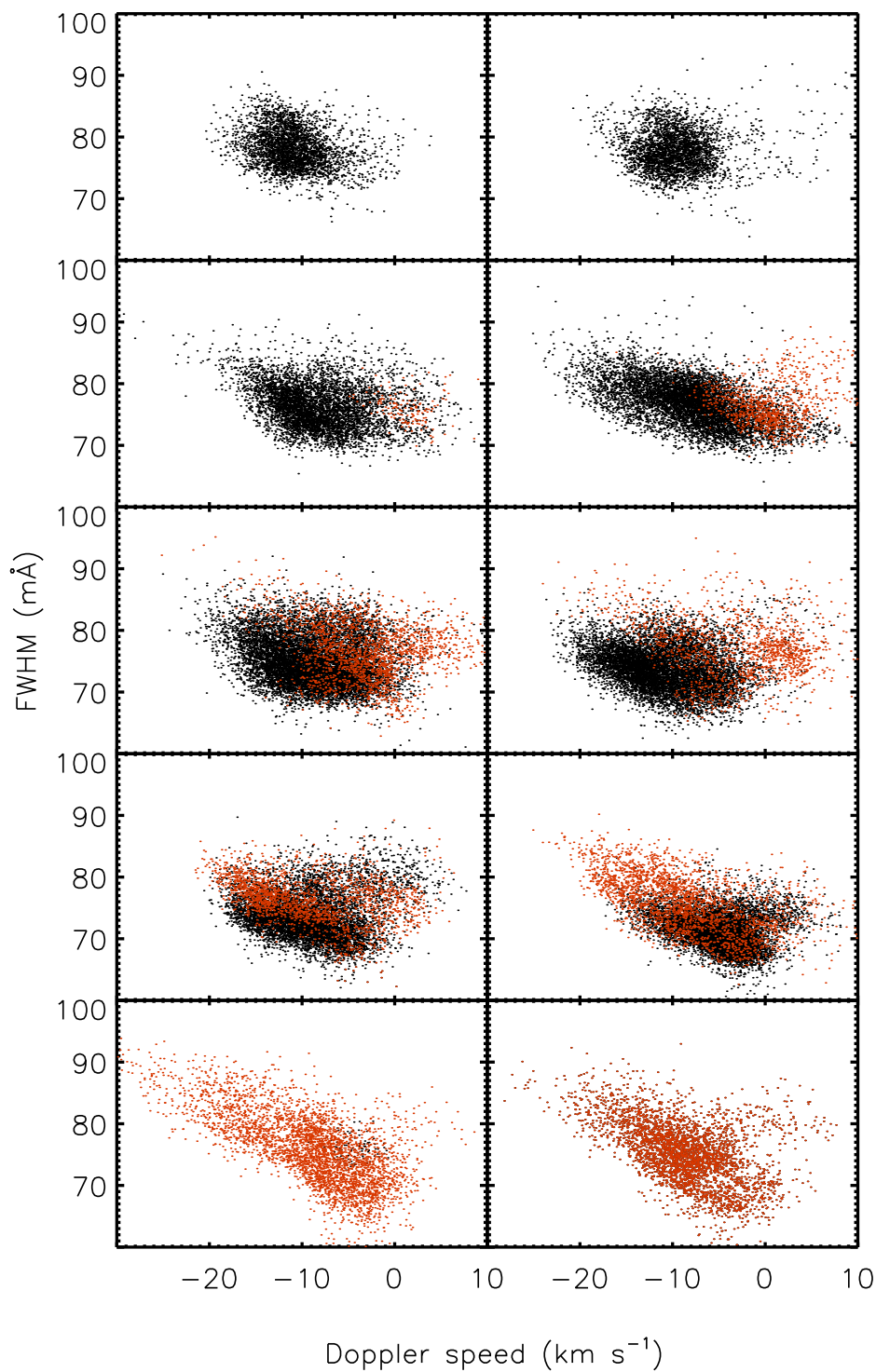}
     \caption{Plots of velocity (centroid shift) versus non-thermal velocity
     (full width half maximum) for the primary Gaussian of
     each of the 10 EIS observations.  
     Black dots represent the western
     outflow region and red dots represent the eastern outflow region.
     Only pixels where the intensity of the secondary component
     is at least 10\% of the intensity of the primary component are plotted.
     The plots follow the same temporal order as
     Figure~\protect\ref{fig:intensity}; left to right and top to bottom.}
     \label{fig:velocity vs width main}
\end{figure}

While the double Gaussian fit appears to represent the \fexii\  spectra well, the
results displayed in Figure~\ref{fig:velocity vs width main} suggest that the
double Gaussian does not describe the outflow completely. In all but the first
two observations there is a correlation between the primary line velocity and
width in at least one of the two outflowing regions.   If the assumption that the
excess line width is due to multiple flow components then this remaining
correlation suggests that a double Gaussian fit does not resolve the flow
components.  The EIS rasters for observations 1 and 2 do not cover the entire
outflowing region, Figures~\ref{fig:intensity} and \ref{fig:intensity secondary}
show that the outflow region is to the north-west of the raster and very likely
extends beyond the raster boundary.  Given that we then do not have data for the
entire outflowing region, it may not be surprising that the same correlation
between velocity and width is not found for these two observations.

\begin{figure}[htp]
     \centering
     \vspace{-1in}
     \includegraphics[width=.80\textwidth]{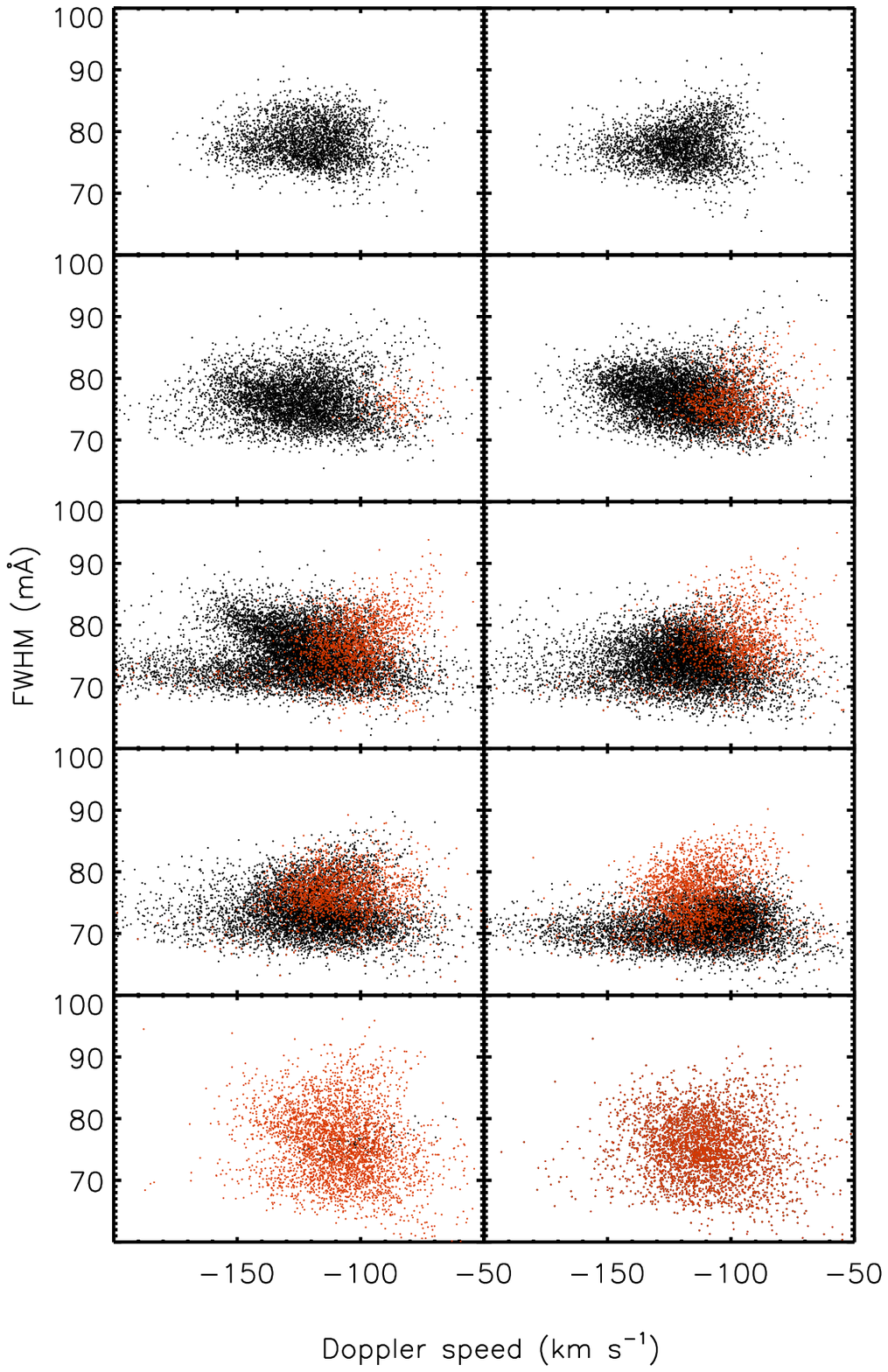}
     \caption{Same as Figure~\protect\ref{fig:velocity vs width main} 
     but for the secondary Gaussian component.}
     \label{fig:velocity vs width minor}
\end{figure}

This correlation that is seen in the primary Gaussian component is not seen for
the secondary component (see Figure~\ref{fig:velocity vs width minor}). One
should be careful on the conclusions drawn from the width in a quantitative 
sense since the width of the primary and secondary components were set as equal
in the fitting algorithm.  However, the fact that there is no correlation between
the velocity and width of the secondary Gaussian component suggests that if the
outflow is indeed comprised of more than two components then the extra,
unresolved, flow components are to be found within the primary component.
The only exception to this is observation 5 where we see a complex relation
between velocity and width, suggesting that we are not fully resolving the
spectral data.

Observations 5, 6, 7, and 8 all show some evidence of regions with high
velocities but low FWHM for the secondary component. The preponderance of these
pixels are found  in areas where the primary outflow speed is relatively low,
away from the loop footpoints. This effect could be due to the direction of the
outflowing material, with the primary and secondary components diverging on
moving away from the footpoints. However, this remains speculation without
knowing the precise topology of the outflowing plasma.

Using a different analysis technique on a different AR, \citet{Pont09a} have
also measured upflows. In contrast to our results, they found upflows throughout
the AR.  While our analysis technique will most readily identify a secondary
component that is not excessively weaker than the primary, we do not believe we
have missed the detection of a secondary component in the AR core. The primary
intensity is a factor $\sim10$ larger in the brightest areas of the AR core
than in the region where we observe the outflow.  If the secondary component had
the same intensity in the AR as found in the outflowing region,  our fitting
technique would find a secondary component of this magnitude should it exist. 
However, we see no indication of a secondary component. Further, the primary
component in the AR does not show the same blue-shift ($\sim10$~km~s$^{-1}$) as
the primary component in the outflowing region.  Should a  secondary component
be present in the AR with similar velocity to that found in the outflowing
region it would be easier to distinguish it from the primary  given the larger
difference in velocities of the two components. In addition, a single Gaussian
fit to the emission in the outflowing region results in excess line widths in
comparison to areas of the observation that show no outflow.  The AR does not
exhibit such excess line widths.

Finally, we note that a double Gaussian fit is a better representation than a
single Gaussian only for the emission from the Fe~{\sc xii} and {\sc xiii}
lines.  Emission from other ions is either too closely affected by 
nearby (in wavelength) emission
lines or too weak in intensity
to determine whether a double Gaussian fit is accurate.

The presence of high velocity outflowing plasma of a duration of several days
leads to the question of where this material is deposited.  It does not appear to
be confined to the loop arcades of the immediate AR; rather it is seen to move
along either open or highly extended field lines.  This indicates significant
mass flow into the corona and, in the case of open field lines, a possible
contributor to the solar wind mass flow.

The velocity that we find for the slow Gaussian component compares well with
models of the slow solar wind by \citet{Wang94a} and \citet{Wang09a}. These
authors determined the outflow velocity from the slow solar wind originating from
small coronal holes in the vicinity of active regions---closely matching the
conditions of the observations analyzed in the present paper.  They predicted
source region outflow velocities of 11.3~km~s$^{-1}$ \citep{Wang94a} and
6.0~km~s$^{-1}$ \citep{Wang09a}.  This is in good agreement with our results
of $\sim10$~km~s$^{-1}$ for the slow moving component of the outflow.
Comparison of 
{\it in situ} measurements of the solar wind are difficult because
of large
coronal holes both proceeding and preceding the active region in question.
It is more likely that any detected solar wind originates in these
coronal holes rather than the outflowing regions analyzed here.

At the 1~arcsec spatial scale of the EIS observations discussed in this paper, we
are unable to spatially
resolve the separate components of the outflowing material.  Thus,
observations of a resolution significantly better than the 1\arcsec\ level will
be required for future instrumentation  in order to  resolve these structures. 
Also, given the temperature dependence of these fast outflows, spectroscopic
information on such a spatial scale is needed to fully determine the nature of
such phenomena.

\acknowledgments

We thank I.\ Ugarte Urra and Y.-M.\ Wang
for helpful discussions and suggestions.

{\it Hinode} is a Japanese mission developed and launched by ISAS/JAXA,
collaborating with NAOJ as domestic partner, and NASA (USA) and STFC (UK) as
international partners. Scientific operation of the {\it Hinode} mission is
conducted by the {\it Hinode} science team organized at ISAS/JAXA. This team
mainly consists of scientists from institutes in the partner countries. Support
for the postlaunch operation is provided by JAXA and NAOJ, STFC, NASA, ESA
(European Space Agency), and NSC (Norway). We are grateful to the {\it Hinode}
team for all their efforts in the design, build, and operation of the mission.

P.\ B., G.\ A.\ D., and P.\ R.\ Y.\ acknowledge support from the NASA {\it
Hinode} program.

\clearpage

\begin{deluxetable}{ccccc}
\tablecaption{AR10978 Observations in 2007\label{tab:obs}}
\tablewidth{0pt}
\tablehead{\colhead{Observation Number} & \colhead{Date} & \colhead{Time (UT)} &
\colhead{Location (arcsec)} & \colhead{EIS Exposure Time (s)}}
\startdata
1  & Dec 09 & 00:12:26 & -663, -191 & 60 \\
2  & Dec 09 & 04:35:39 & -631, -190 & 60 \\
3  & Dec 10 & 00:19:27 & -447, -161 & 40 \\
4  & Dec 11 & 00:24:16 & -178, -144 & 60 \\
5  & Dec 11 & 10:25:42 & -150, -141 & 40 \\
6  & Dec 12 & 03:26:43 & 16,   -133 & 40 \\
7  & Dec 12 & 11:43:36 & 91,   -128 & 40 \\
8  & Dec 13 & 12:18:42 & 347,  -134 & 40 \\
9  & Dec 15 & 00:13:49 & 620,  -142 & 40 \\
10 & Dec 15 & 18:15:49 & 737,  -142 & 40 
\enddata
\tablecomments{Times given are the start times of each observation.
Locations are the center of the raster.  Exposure times are for each position
in the raster.}
\end{deluxetable}


\end{document}